\documentclass[twocolumn,twocolappendix]{aastex631}
\usepackage[mathlines]{lineno}
\usepackage{xspace}
\usepackage{amsmath}
\usepackage{multirow}

\newcommand{\photoz}{photo-$z$\xspace}
\newcommand{\specz}{spec-$z$\xspace}
\newcommand{\rp}{$r_{\rm p}$\xspace}

\received{xxx, xxx}
\revised{xxx, xxx}
\accepted{xxx}
\submitjournal{ApJ}
\shorttitle{MMG evolution}
\shortauthors{Li et al.}

\graphicspath{{./}{figures/}}

\begin{document}

\title{Evolution of the Physical Properties of the Most Massive Galaxies in Clusters and their Protohalos}

\correspondingauthor{Qingyang Li, Xiaohu Yang}
\email{qingyli@sjtu.edu.cn, xyang@sjtu.edu.cn}

\author[0000-0003-0771-1350]{Qingyang Li}
\affiliation{State Key Laboratory of Dark Matter Physics, Tsung-Dao Lee Institute \& School of Physics and Astronomy, Shanghai Jiao Tong University, Shanghai 200240, China}
\affiliation{Shanghai Key Laboratory for Particle Physics and Cosmology, Shanghai Jiao Tong University, Shanghai 200240, China}

\author[0000-0003-3997-4606]{Xiaohu Yang}
\affiliation{State Key Laboratory of Dark Matter Physics, Tsung-Dao Lee Institute \& School of Physics and Astronomy, Shanghai Jiao Tong University, Shanghai 200240, China}
\affiliation{Shanghai Key Laboratory for Particle Physics and Cosmology, Shanghai Jiao Tong University, Shanghai 200240, China}

\author{Antonios Katsianis}
\affiliation{School of Physics and Astronomy, Sun Yat-sen University, Zhuhai Campus, Zhuhai 519082, China}

\author{Paola Popesso}
\affiliation{European Southern Observatory, Karl Schwarzschildstrasse 2, 85748, Garching bei Munchen, Germany}
\affiliation{Excellence Cluster ORIGINS, Boltzmannstr. 2, D-85748 Garching bei Munchen, Germany}

\author[0000-0003-2130-2537]{Ilaria Marini}
\affiliation{European Southern Observatory, Karl Schwarzschildstrasse 2, 85748, Garching bei Munchen, Germany}

\author[0000-0002-7928-416X]{Y. Sophia Dai}
\affiliation{Chinese Academy of Sciences South America Center for Astronomy (CASSACA), National Astronomical Observatories of China, Chinese Academy of Sciences, 20A Datun Road, Beijing, China}

\author[0000-0002-4718-3428]{Chengze Liu}
\affiliation{State Key Laboratory of Dark Matter Physics, Tsung-Dao Lee Institute \& School of Physics and Astronomy, Shanghai Jiao Tong University, Shanghai 200240, China}
\affiliation{Shanghai Key Laboratory for Particle Physics and Cosmology, Shanghai Jiao Tong University, Shanghai 200240, China}

\author[0000-0002-4534-3125]{Yipeng Jing}
\affiliation{State Key Laboratory of Dark Matter Physics, Tsung-Dao Lee Institute \& School of Physics and Astronomy, Shanghai Jiao Tong University, Shanghai 200240, China}
\affiliation{Shanghai Key Laboratory for Particle Physics and Cosmology, Shanghai Jiao Tong University, Shanghai 200240, China}

\author[0000-0001-6511-8745]{Jia-Sheng Huang}
\affiliation{Chinese Academy of Sciences South America Center for Astronomy (CASSACA), National Astronomical Observatories of China, Chinese Academy of Sciences, 20A Datun Road, Beijing, China}

\author[0000-0002-7712-7857]{Marcin Sawicki}
\affiliation{Department of Astronomy and Physics and Institute for Computational Astrophysics, Saint Mary’s University, 923 Robie Street, Halifax, NS, B3H 3C3, Canada}

\begin{abstract}
We investigated the evolution of the physical properties of the brightest galaxies in clusters and their protohalos from $z = 4$ to $z = 0$. Galaxy clusters and groups are identified using a halo-based group finder applied to the COSMOS2020 galaxy catalog. We construct evolution chains from low redshift clusters to higher redshift groups via the abundance matching method. The region of protohalos corresponding to clusters is defined on the basis of a characteristic radius. Our analysis encompasses a wide range of physical properties, including stellar mass, luminosity, star formation rate (SFR), specific star formation rate (sSFR), color ($g - r$), and stellar age. 
The evolution trends of the most massive galaxies (MMGs) in higher redshift groups and their corresponding protohalos are generally consistent.
The stellar mass of MMGs shows an increasing trend across the entire redshift range. By considering the stellar mass growth as in-situ and ex-situ components, we find that in-situ star formation is efficient at $z \sim 2$, while ex-situ accretion becomes the primary growth channel at later times. 
At $z \gtrsim 2$, MMGs undergo an intense star formation phase of approximately $10^{2}\ \rm M_{\odot}yr^{-1}$, but are generally quenched at lower redshifts. Stellar age analysis suggests that most stars in MMGs formed at $z > 2$. 
Our results present a coherent picture of MMG evolution across cosmic epochs, which is broadly consistent with the current theoretical framework of galaxy formation and evolution. Moreover, our work provides an intriguing way to trace galaxy evolution through the construction of cluster evolutionary chains in observations.
\end{abstract}

\keywords{Galaxy groups(597) --- Galaxy clusters(584) --- Galaxy dark matter halos(1880) --- Protoclusters(1297) --- Galaxy evolution(594) --- Galaxies(573) --- Dark matter(353)}

\section{Introduction} \label{sec:intro}

The most massive (or luminous) galaxies (MMGs) in clusters/groups, also known as the brightest cluster galaxies (BCGs), are a unique class of objects in the Universe \citep{Lin2004}. Their exceptional special properties, such as large mass and extended spatial extent, are distinct from the general population of galaxy within clusters, making them ideal objects for studies of galaxy formation \citep{DeLucia2007} and classical cosmological tests \citep{Collins1998, Harvey2019}. Moreover, since MMGs reside at the deepest points of the gravitational potential wells within the most massive collapsed structures of the Universe (often coinciding with the peaks of X-ray emission), they are likely influenced by a combination of simultaneous physical processes, including galaxy mergers, cluster cooling flows, AGN feedback, and star formation. Therefore, MMGs are considered to be intimately connected with the development of galaxy clusters and serve as crucial tools for investigating the key processes influencing galaxy evolution.

The evolution of MMGs has been widely investigated in numerous studies involving various physical properties, such as stellar mass \citep[e.g.,][]{Cooke2018, Cerulo2019}, luminosity \citep{Whiley2008, Stott2008, Chu2022}, star formation rate (SFR)/specific star formation rate (sSFR) \citep[e.g.,][]{Fraser2014, McDonald2016, Cooke2018, Orellana2022}, color \citep[e.g.,][]{Tonini2012,Cerulo2019}, age \citep[e.g.,][]{Edward2023, Gozalias2024}. Existing literature primarily examines the objects in the nearby Universe with an accurate measurement of the properties of MMGs through spectroscopic data. For example, \citet{Cerulo2019} studied the evolution of color, stellar mass, and star formation activity in nearby MMGs in a sample of clusters at $0.05 < z < 0.35$ from the Sloan Digital Sky Survey (SDSS) and the Wide Infrared Survey Explorer (\emph{WISE}). They did not find significant stellar mass growth between $z = 0.35$ and $z = 0.05$, although there is a debate among different authors regarding the growth rate of stellar mass in recent times \citep[e.g.,][]{Stott2008, Lidman2012, Lin2013, Oliva2014, Zhang2016, Lavoie2016, Orellana2022}. In addition, \citet{Orellana2022} investigated the stellar mass and star formation activities through a large sample of $\sim 56000$ MMGs from SDSS and \emph{WISE} at $0.05 < z < 0.42$.

However, high-redshift MMGs are difficult to observe. Only a limited number of studies have been conducted based on the galaxy samples from deep survey fields such as the COSMOS field, typically reaching a redshift up to $\sim$2.0 \citep[e.g.,][]{Wen2011, Webb2015, Bellstedt2016, Bonaventura2017, Zhao2017}. \citet{Cooke2019} examined the role of the environment in the in situ stellar mass growth of the progenitors of the MMGs in the present-day Universe in COSMOS from $z \sim 3$ to the present. \citet{Chu2021} investigated the properties of MMGs, including S$\rm \acute{e}$rsic index, effective radius, position angle of the major axis, and surface brightness up to redshift 1.8 based on Hubble Space Telescope (\emph{HST}) data. Recently, \citet{Gozalias2024} used the MMGs selected from the X-ray group catalog within the COSMOS field to analyze the evolution of stellar properties at $0.08 < z < 1.30$.  

At $z \gtrsim 2$, the environment of MMGs becomes complex, as galaxy clusters are not yet virialized and are still in the process of formation, commonly referred to as protoclusters \citep[see][for recent reviews]{Kravtsov2012, Overzier2016}. However, identifying protoclusters in observation is challenging, as it typically requires deep observations with a long exposure time to detect distant and faint galaxies. A convenient and universal way to determine protoclusters is to trace the high density star-forming regions guided by the presence of particular objects, such as H$\alpha$ emitters and Ly$\alpha$ emitters \citep[e.g.,][]{Katsianis2017, Cai2019, Hu2021}. In addition, since the groups identified by the group finder are a theoretically virialized structure, high-redshift groups are thought to be the cores of protoclusters, providing an efficient way to identify protoclusters \citep{Ando2022}. For example, \citet{Li2022} identified more than 1,000 protocluster candidates based on high-redshift groups at $2 < z < 6$ using the CLAUDS and HSC-SSP joint deep survey data. In this work, we will extend the investigation of the physical properties of MMGs from galaxy groups to protocluster environments, reaching redshift up to $\sim 4$.  

In addition to observations, the evolutionary history of MMG can be effectively explored using semi-analytic models \citep[e.g.,][]{DeLucia2007,Tonini2012} or hydrodynamical simulations \citep{Ragone2018,Remus2023,Fukushima2023}. The advantage of using simulations is that the physical properties and environments of high-redshift MMGs can be directly extracted. Moreover, both MMGs and their environments can be continuously traced over cosmic time. In contrast to protoclusters in observation, simulations can track the progenitors of groups, usually referred to as protohalos from the point of view of dark matter halos, which include all the matter that will eventually collapse into a group at $z = 0$. In addition, the characterized radius of protoclusters has been developed and studied from simulations in order to quantify the growth of the size and mass of the protoclusters \citep[e.g.,][]{Chiang2013, Muldrew2015}, thus providing theoretical guidance to identify protoclusters in observations. For example, \citet{Wang2023} used a double power law function to describe the protohalo size history based on the ELUCID simulation \citep{Wang2014}. 

As spectroscopic observations of high-redshift galaxies are expensive and time-consuming, photometric observation provides an efficient alternative. One of the most well-known photometric survey fields is the COSMOS field, which has been extensively observed by multiple deep surveys, including \emph{HST}, HSC-SSP, \emph{Spitzer}. The availability of multiband photometry makes the estimation of redshift and physical properties of galaxies through SED fitting more accurate than only using a few photometry bands. In this work, we use the COSMOS2020 photometric catalog, which combines multiple datasets from various surveys in the COSMOS field, to investigate the evolution of MMGs in their counterpart groups and protohalos \footnote{As the range of a protocluster is defined with a protohalo size, we refer to protoclusters as protohalos in this study.}, extending the property analysis to $z \sim 4$. We compare two types of MMGs: first and second MMGs\footnote{The rank of MMGs is determined from the magnitude of stellar mass.}. The different formation histories, where the first MMGs reside at the cluster centers, while the second MMGs experience an infall process, make them ideal objects to contrast the effects of environments. The growth history of stellar mass for MMGs is explored based on the evolution of stellar mass and SFR. Clusters/groups are identified using a halo-based group finder \citep{Yang2021}. Instead of directly comparing MMGs at fixed redshift bins, we construct an evolutionary sequence of groups based on the abundance matching method to study the physical properties of MMGs over cosmic time. In parallel, we examine the evolution of MMGs within protohalo environments.

In Section~\ref{sec:data}, we introduce the galaxy sample selected from the COSMOS2020 catalog. We introduce the method to determine groups and protohalos in Section~\ref{sec:method}. We show the different physical properties of the evolution of MMGs in Section~\ref{sec:results}. We discuss the main driver of stellar mass growth in Section~\ref{sec:two_phase} and compare the methods of tracing MMGs in Section~\ref{sec:disc}. We summarize our conclusions in Section~\ref{sec:summary}.
Throughout the paper, we adopt a $\Lambda$CDM cosmology with parameters that are consistent with the \emph{Planck} 2018 results \citep{Planck2020}: $\Omega_\mathrm{m} = 0.315$, $\Omega_{\Lambda} = 0.685$, $n_{\rm s} = 0.965$, $h = H_0/(100\ \rm km\ s^{-1}\ Mpc^{-1)}=0.674$ and $\sigma_8=0.811$.



\section{Data} \label{sec:data}

\subsection{COSMOS2020}

We construct the galaxy sample mainly on the basis of the COSMOS2020 galaxy catalog \citep{Weaver2022}, which specifically focuses on the 2 $\mathrm{deg}^2$ of the COSMOS field. This catalog provides approximately 966,000 galaxies that are measured with all available UV-optical-IR data. The primary data sets include the CFHT Large Area $U$ band Deep Survey \citep[CLAUDS, ][]{Sawicki2019}, HST/ACS F814W photometry, the second public data release of the Hyper Suprime-Cam Subaru Strategic Program \citep[HSC-SSP, ][]{Aihara2019}, Subaru Suprime-Cam data, the fourth data release of the UltraVISTA survey \citep{McCracken2012, Moneti2023}, and Spitzer/IRAC channels \citep{Moneti2022}. In total, more than 15 bands from near-UV to mid-infrared have reached a depth of $\sim 25$ mag at 3$\sigma$ depth, measured in $2''$ diameter apertures at least. 
The catalog provides two ways to extract photometry information: a traditional aperture photometric method \citep[CLASSIC, ][]{Laigle2016, Hsieh2012} and a new profile-fitting photometric extraction tool \citep[THE FARMER, ][]{Weaver2023}. Photometric redshifts (\photoz), as well as physical properties of galaxies (e.g. stellar mass and SFR), are computed using both \textsc{LePhare} \citep{Arnouts2002, Ilbert2006} and \textsc{EAZY} \citep{Brammer2008} for both the CLASSIC and THE FARMER catalogs. 

In this work, we use the CLASSIC catalogs with the \photoz and the physical properties computed from \textsc{LePhare}. Here we briefly introduce the implementation of the SED fitting in \textsc{LePhare}. More details can be found in \citet{Weaver2022} and \citet{Laigle2016}. Stellar templates include the library of \citet{Pickles1998}, the white dwarf templates \citep{Bohlin1995}, and dwarf templates from \citet{Chabrier2000}, \citet{Baraffe2015} and \citet{Morley2012, Morley2014}. Galaxies templates include elliptical and spiral galaxy models from \citet{Polletta2007}, which were interpolated into 19 templates to increase the resolution, and 12 blue star-forming galaxy models from \citet{Bruzual2003}. In addition, two additional templates from \citet{Bruzual2003} with exponentially declining SFR were added to improve the photo-$z$ of quiescent galaxies. Extinction is a free parameter with reddening $E(B - V) \leq 0.5$. The predicted fluxes for the templates are computed using a redshift grid with a step of 0.01 and a maximum redshift of 10. Emission lines are considered using the relation between the UV luminosity and [O\,\textsc{ii}] emission-line flux, as well as fixed ratios between dust-corrected emission lines following \citet{Ilbert2009}. \textsc{LePhare} also includes a set of AGN and quasar templates \citep{Salvato2009}. As for the fitting process, an initial run of \textsc{LePhare} is used to fit the galaxies with spectroscopic redshift (\specz) to optimize the absolute calibration in each band. The running of \textsc{LePhare} produces important quantities such as \photoz, absolute magnitude, and the physical properties of galaxies.  

In the next section, we illustrate the selection of our galaxy sample based on the CLASSIC catalog, supplemented by additional low-redshift galaxies drawn from other catalogs.

\subsection{Galaxy sample}

Our galaxy sample first rejects the objects with conservative combined mask flags from HSC, UltraVISTA, and IRAC, to avoid the light influence from bright stars. Sources are separated into galaxies, stars, and AGNs in \textsc{LePhare} by combining the morphological and SED criteria, where we only select galaxies. In this work, we focus on the HSC-SSP $i$-band galaxies, since the $i$ band is the priority observed in the HSC-SSP observation, where the $i$ band has the smallest seeing. We choose galaxies with an apparent magnitude limit $i < 25$ (in $2''$ aperture). The uncertainty of magnitude under this condition is less than 0.05. We selected galaxies in the \photoz range $0 < z < 5$ and the $i$-band absolute magnitude measured by \textsc{LePhare} larger than $-$10.86. The number of galaxies after these selections is 171,541. 

\begin{figure}
    \centering
    \includegraphics[width=0.48\textwidth]{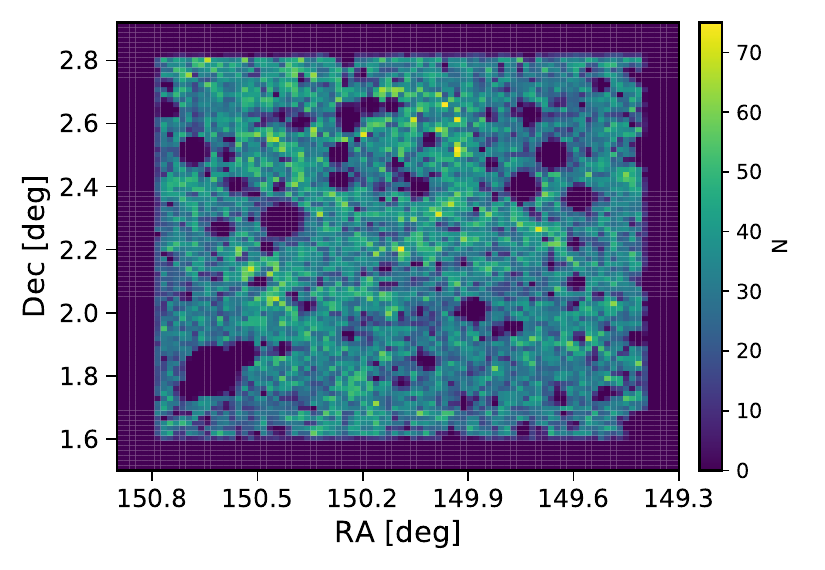}
    \caption{The sky coverage of the selected galaxies in the COSMOS field. The galaxy number count in each pixel with an area of about $2.8 \times 10^{-4}\ \rm deg^2$ is coded with the color bar. The empty circles inside the coverage of the galaxies correspond to the masked areas.}
    \label{fig:map}
\end{figure}

\begin{figure}
    \centering
    \includegraphics[width=0.48\textwidth]{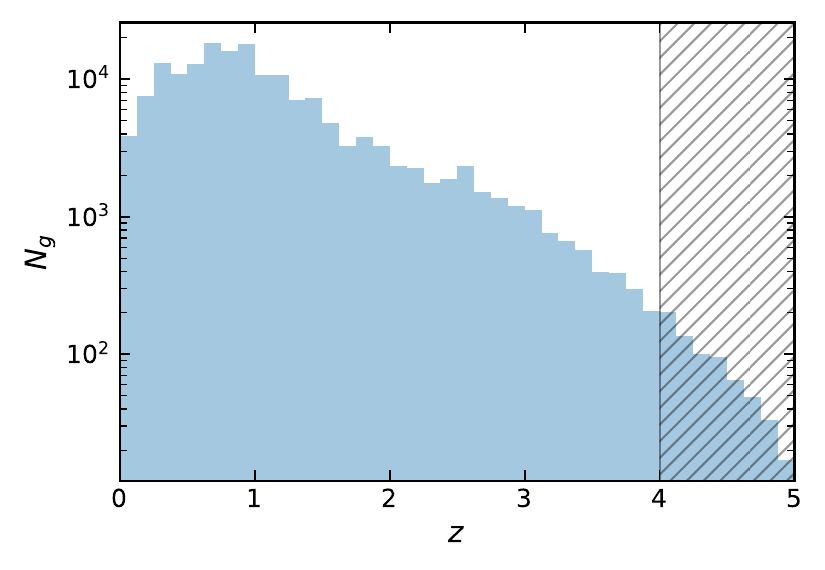}
    \caption{The number histogram of the selected galaxies as a function of \photoz. The galaxies at $4 < z < 5$ marked by the shadow region are not used for MMG evolution analysis.}
    \label{fig:ngal}
\end{figure}

Then, we replenish the low-redshift galaxies that may be excluded due to the mask effect. These galaxies are chosen from the DESI DR9 galaxy catalog (an updated version of \citealt{Yang2021}) and the 2020 version of the Siena Galaxy Atlas (SGA) catalog \citep{Moustakas2023}, which focuses on the nearby galaxies. The DESI DR9 galaxy catalog, based on the DESI Legacy Imaging Surveys, has a $z$-band magnitude limit of 21 and includes galaxies with redshifts below 1. From this catalog, we select bright galaxies with $z$-band magnitudes brighter than 18. These two galaxy catalogs result in 516 additional galaxies in the corresponding COSMOS region. We match this galaxy sample with the original COSMOS2020 galaxy sample and exclude galaxies without a measurement of \textsc{LePhare}, as well as duplicate entries already included in the selected COSMOS2020 sample. This finally provides 378 additional galaxies incorporated into the selected COSMOS2020 galaxy sample. In total, we have 171,799 galaxies that span an area of 1.50 $\rm deg^2$ with the selection criteria. In Figure~\ref{fig:map}, we show the footprint of selected galaxies from COSMOS2020 on the sky. The \photoz distribution of selected galaxies is presented in Figure~\ref{fig:ngal}. 

The performance of \photoz at the $i$ band has been specifically investigated in \citet{Weaver2022} with a comparison of the \specz. The quantities used to evaluate the performance of \photoz include the bias $b$, the normalized median absolute deviation (NMAD), and the outlier rate $\eta$. The $b$ is defined as the median difference between \photoz and \specz. The NMAD, also regarded as a \photoz uncertainty, is defined as:
\begin{equation}
    \sigma_{\rm NMAD} = 1.48 \times \mathrm{med}\left(\frac{|\Delta z - \mathrm{med}(\Delta z)|}{1 + {z_{\rm spec}}}\right),
\end{equation}
where $\Delta z$ is the difference between \photoz and \specz.
The $\eta$ is defined as the fraction of galaxies with $|\Delta z| > 0.15(1 + z_{\rm spec})$. The values of $b$ in three magnitude bins $17 < i < 22.5$, $22.5 < i < 24.0$, and $24.0 < i < 25.0$ are $-0.003$, $-0.004$ and $-0.002$ with uncertainty of about 0.008, 0.015 and 0.024, while the corresponding $\eta$ are 0.6$\%$, 3.2$\%$ and 6.3$\%$ \citep[as shown Figure 13 in][]{Weaver2022}. Overall, there is almost no bias at different magnitude bins, despite a slightly high $\eta$ in the faint range. 
We determine the \photoz error of each galaxy based on the redshift evolution curve of the median of \photoz 1$\sigma$ uncertainties \citep[as shown the Figure 17 in][]{Weaver2022}. The distribution of \photoz uncertainty is considered in two magnitude bins $17 < i \leq 24$ and $24 < i \leq 25$ that contain 74,003 and 97,796 galaxies, respectively. In general, the average \photoz uncertainties in the two magnitude bins are about 0.028 and 0.083.
Next, we introduce the methods used to determine galaxy clusters/groups and their corresponding protohalos, as well as the approach to construct the group evolution chain from high redshift to low redshift for clusters. 

\section{Method} \label{sec:method}

We used the halo-based group finder presented in \citet{Yang2007} to identify clusters/groups in the galaxy sample. The progenitors of low redshift clusters/groups are traced using the abundance matching method, thereby constructing the evolutionary sequence of clusters. Finally, the corresponding protohalos for each cluster are defined on the basis of two key factors: the cores of the protohalos and the characteristic size of the protohalos.

\subsection{The halo-based group finder} \label{sec:gfinder}

The group finder that we used in this study is a halo-based method developed in \citet{Yang2005,Yang2007}. The galaxy-halo connections have been extensively studied in theory (e.g., halo occupation distribution, stellar to halo mass ratios), which provide the foundation of this group finder. In its first version, the group finder was only applicable to galaxies with \specz, and hence has been applied to the 2dFGRS and SDSS data to search for groups at low redshifts \citep{Yang2005,Yang2007}. Now, the group finder has been improved to be simultaneously suitable for both spectroscopic and photometric redshifts. With an application to mock galaxy redshift samples, the group finder has been verified to be stable and reliable in constructing group catalogs for galaxy samples with photometric redshift error $\sim 3$\% \citep{Yang2021}. The updated version of group finder has been successfully applied to the DESI Legacy Imaging Surveys covering a large sky area \citep{Yang2021} and the CLAUDS and HSC-SSP joint deep surveys in a wide redshift range \citep[$0 < z < 6$, ][]{Li2022}. Here we give a brief description of the main steps of this method \citep[more details can be found in][]{Yang2021, Li2022}. 

The group finder begins with the assumption that each galaxy is a group candidate. We bin our sample galaxies with a redshift interval of $\Delta z$ = 0.4. Then, each tentative group is assigned a halo mass $M_{h}$ based on the mass-to-light ratio using interpolation techniques. The mass-to-light ratios of the groups in each redshift bin are determined with the cumulative halo mass functions \citep{Sheth2001} and the group luminosity functions using the abundance matching method \citep{Yang2007}. The luminosity of each galaxy is obtained by the formula: 
\begin{equation} \label{eq:lum}
    \log (L/h^{-2}\mathrm{L_{\odot}}) = 0.4 \times (4.52 - M_i)\,,
\end{equation}
where 4.52 is the $i$-band absolute magnitude of the Sun \citep{Willmer2018} and $M_i$ is the $i$-band absolute magnitude of each galaxy given by \textsc{LePhare}.

With a halo mass, each group can have a halo radius and velocity dispersion along the line-of-sight. The halo radius is defined as 180 times the average matter density of the universe. The line-of-sight velocity dispersion of a dark matter halo, $\sigma_{180}$, is obtained using the fitting function of \citet{van2004} with slight modifications to be suitable to $\Lambda$CDM cosmology with other $\Omega_m$ values. The group membership updates begin from the most massive one by taking the luminosity-weighted group center as the halo center and assuming that the distribution of member galaxies in phase space follows that of the dark matter particles. The probability of a galaxy being a member galaxy can be expressed as:
\begin{equation}
    P_\mathrm{M}(R, z_\mathrm{d}) = \frac{H_0}{c} \frac{\Sigma(R)}{\bar{\rho}} p(z_\mathrm{d}),
\end{equation}
where $R$ is the projected distance from the group center, $z_\mathrm{d} = z - z_{\rm group}$, $c$ is the velocity of light, $\Sigma (R)$ is the projected surface density for a NFW halo \citep{Navarro1997}, $p(z_\mathrm{d})$ is a Gaussian function form to describe the redshift distribution of galaxies within the halo. We combine the contribution of \photoz error $\sigma_{\rm NAMD}$ and $\sigma_{180}$, that is, $\sigma= \sqrt{ \sigma_{180}^2+ c^2\sigma_{\rm NAMD}^2} $, which is used to describe the dispersion in $p(z_\mathrm{d})$. Galaxies are assigned to a candidate group with a judgement between $P_\mathrm{M}(R,z_\mathrm{d})$ and $\mathrm{B}\sigma_{180}/\sigma$. Here, the background value $B$ is independent of halo mass and perceptively quantifies the threshold of the redshift space density contrast of groups \citep{Yang2005}. The ratio $\sigma_{180}/\sigma$ is used to account for the decrease in density contrast caused by the photometric redshift error. If $P_\mathrm{M}(R,z_\mathrm{d}) \ge \mathrm{B}\sigma_{180}/\sigma$, the galaxy will be assigned to the group. We adopt the theoretically gauged parameter 10 as the value $B$ during group finding. Decreasing this background value may slightly increase the richness of the groups. 

After assigning all the galaxies into groups, we update the group centers and luminosities, recalculate the halo information and find member galaxies again. The iterative stops until there are no more changes for the group memberships. Finally, we start from the beginning to make another iteration, aimed at the convergence of mass-to-light ratios, which normally need 3 to 4 iterations.

We obtain a total of 102,831 groups after applying the halo-based group finder to the selected galaxy sample\footnote{We use the absolute rest-frame AB magnitude of $i$ band in the group finder.}. There are 11,842 and 997 groups containing at least 3 and 10 member galaxies, respectively. In the group catalog, 8 clusters/groups have a halo mass of $M_h > 10^{14}\  \rm M_{\odot}$ within the redshift range $0 < z < 0.4$. The concrete information about these 8 groups is shown in Table~\ref{tab:basicgroup}. For convenience, we call these 8 groups the \emph{reference} clusters in the following. We then trace the progenitors of these reference clusters and determine their protohalos. 

\begin{table}
    \centering
    \caption{The information of reference clusters.}
    \label{tab:basicgroup}
    \begin{tabular}{c c c c c c c }
    \hline \hline
     ID & $N_{\rm mem}$ & RA & Dec & $z$ & $M_h$ & $L$ \\
     \hline
     (1) & (2) & (3) & (4) & (5) & (6) & (7) \\
     \hline
      1 &  3093  & 149.9288 & 2.0987 & 0.1112 & $10^{15.38}$ & $10^{12.96}$  \\
      2 &  248  & 150.1997 & 165.4700 & 0.2204 & $10^{14.48}$ & $10^{12.39}$  \\
      3 &  179  & 150.6330 & 2.7438 & 0.2071 & $10^{14.05}$ & $10^{11.98}$  \\
      5 &  153  & 150.4424 & 2.0571 & 0.3200 & $10^{14.19}$ & $10^{12.12}$  \\
      7 &  121  & 149.4281 & 2.5701 & 0.3737 & $10^{14.36}$ & $10^{12.31}$  \\
      8 &  120  & 150.1831 & 1.7711 & 0.3399 & $10^{14.08}$ & $10^{12.00}$  \\
      13 &  97  & 149.8119 & 2.1476 & 0.3535 & $10^{14.14}$ & $10^{12.05}$  \\
      16 &  93  & 149.9630 & 1.6807 & 0.3715 & $10^{14.02}$ & $10^{11.97}$  \\   
    \hline
\end{tabular}
\tablecomments{Column (1) lists the group ID in our group catalog. Column (2), (3), (4), (5), (6), and (7) list the number of member galaxies, RA, Dec, redshift, halo mass, and total luminosity, respectively. The units of halo mass and luminosity are $\rm M_{\odot}$ and $\rm L_{\odot}$.}
\end{table}

\subsection{The progenitors of reference clusters} \label{sec:groups}

\begin{figure*}
    \centering
    \includegraphics[width=0.98\textwidth]{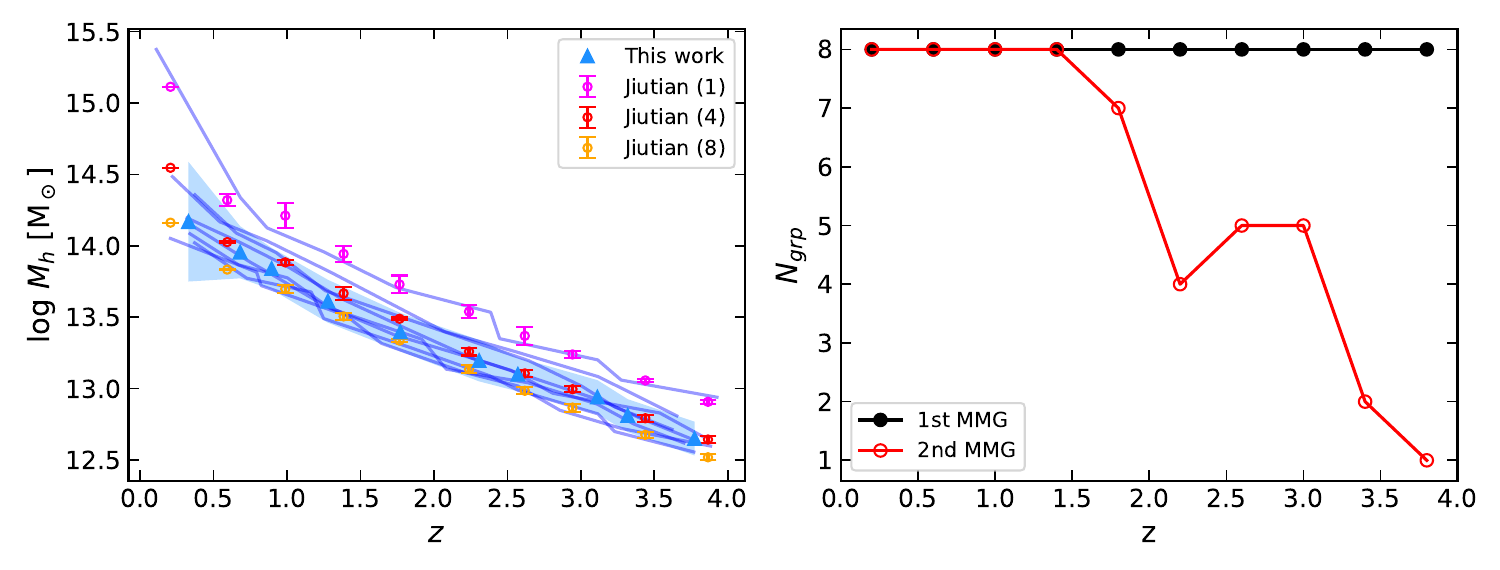}
    \caption{Left: the halo mass evolution of the progenitors of reference clusters. Each blue line represents the evolution of individual reference clusters. The blue triangles show the median halo mass with shaded area indicating $1\sigma$ error. The magenta, red, and orange points show the median halo mass corresponding to the first, fourth, and eighth most massive halos within the redshift bin $0 < z < 0.4$ in the same level of CHMF. The halo sample at each redshift was obtained from 200 sub-boxes carved out of the 1 $h^{-1}\mathrm{Gpc}$ cubic box of Jiutian simulation. The error bars indicate $1\sigma$ uncertainties. Right: the number of matched groups possessing the first MMG (solid black points) and the second MMG (hollow red points).}
    \label{fig:sta_group}
\end{figure*}

While the progenitors of clusters can be conveniently traced from simulations, directly tracing the evolutionary history of an individual cluster using solely observations is difficult. Through simulations, a comprehensive grasp of the evolution of dark matter halos and their connection to galaxies has been achieved, offering an impressive method to develop a pseudo-evolutionary sequence of groups from observational data. Based on the hierarchical halo formation framework, the progenitors of clusters found in observation can be divided into two types: (1) the virialized low-redshift clusters, which represent the final products of halo formation, and (2) high-redshift groups, which correspond to the main branches in the merger history of halos. The key challenge lies in linking these virialized low-redshift clusters to their high-redshift progenitors. To address this, we adopt the abundance matching method. 

The groups in different redshift bins are connected on the basis of the abundance matching for the halo mass function. Specifically, we first compute the cumulative halo mass function (CHMF) in redshift bins of width $\Delta z = 0.4$, consistent with the redshift interval used in our group finder. 
Within each redshift bin, groups are ranked by halo mass. The low redshift groups are matched to the high redshift groups with the same level of CHMF. The groups with the corresponding rank sequence at higher redshifts are determined as the progenitors of the low-redshift clusters/groups. This approach ensures that more massive groups are linked to more massive progenitors with constant comoving density. As the number of relatively massive groups at $z > 4$ is small, we trace the progenitors of the groups only up to $z = 4$, which means the number of redshift bins used to identify progenitors is 10. In the left panel of Figure~\ref{fig:sta_group}, we present the halo mass evolution of the matched progenitors of reference clusters, which show an increasing trend with redshifts decreased as expected. In addition, we compare with a halo sample selected from Jiutian simulation \citep{Han2025} to show the effect of cosmic variance on the abundance matching method, especially when the area of the COSMOS field is small. We choose the run in periodic boxes of 1 $h^{-1}\rm Gpc$ with $6144^3$ particles and select the snapshots corresponding to the mean redshift of each redshift interval used to match progenitors. Then, we carve out 200 sub-boxes at each snapshot, of which the volume is equal to the size of the redshift interval. For each sub-box, we select top-ranked halos by mass. 
We compute the median halo mass at each redshift for the corresponding matched halos of the first, fourth, and eighth most massive halos within redshift bin $0 < z < 0.4$ in the same level of CHMF, which are shown with the magenta, red, and orange points in the left panel of Figure~\ref{fig:sta_group}. The halo masses of the progenitors of reference clusters follow the evolution of the fourth most massive halos and lie between the trends of the first and eighth most massive halos across the redshift range, suggesting that the inferred evolutionary trend of reference clusters is less affected by cosmic variance.

As we determine the evolutionary progression of groups, we can also identify the progenitors of MMGs by choosing the most massive galaxies in their respective progenitor groups. In the right panel of Figure~\ref{fig:sta_group}, we examine the situations of the first MMGs and second MMGs in the matched groups. At $z < 1.5$, each group linked to the reference cluster contains both the first MMG and the second MMG. However, the number of groups possessing second MMG begins to decrease at $z > 1.5$, indicating that some groups contain only a single galaxy.

\subsection{Protohalos} \label{sec:protohalos}

A protohalo is the collection of all progenitor halos at a given $z > 0$ that would end up in a common descendant halo at $z = 0$. In the previous section, we determined the progenitors of our eight reference clusters, which are regarded as the main progenitors, i.e., the main branch of the halo formation history. Thus, these groups can be considered to be the cores of protohalos. The other progenitors are included within a characteristic radius that defines the full extent of the protohalo. The radius of protohalos was larger in the past. This reflects the expected behavior of collapsing structures in an expanding universe, where earlier (higher-redshift) structures were more extended. We use a double power-law formula developed by \citet{Wang2023} to trace the protohalo size history. The size of a protohalo is defined as the mass-weighted radius of all progenitor subhalos relative to the protohalo center. More details can be found in \citet{Wang2023}. 
The formula used to describe the size of protohalos, $R_{\rm ph}(z)$, as a function of redshift is expressed as:
\begin{equation}
    R_{\rm ph}(z) = \frac{2R_c(z/2)^{\alpha}}{1 + (z/2)^{\alpha - \beta}},
\end{equation}
where $\alpha$ and $\beta$ are the late-time and early-time slope, respectively. The values are determined through the relation to the mass of the descendant halo. $\beta$ has a median value of $-0.2$ independent of the descendant halo mass. $\alpha$ has a slight dependence on the descendant halo mass, expressed as:
\begin{equation}
    \alpha = -0.07 \log M_{h0} + 1.99,
\end{equation}
where $\log M_{h0}$ is the descendant halo mass in unit of $h^{-1}M_{\odot}$ at redshift $z=0$. $R_c$ is the amplitude of the protohalo size history, following the formula:
\begin{equation}
    \log (R_c/h^{-1}\mathrm{Mpc}) = (0.31m - 4.80)r + 0.28m - 3.55,
\end{equation}
where $m$ is the logarithm halo mass of the descendant group $\log M_{h0}$ in unit of $h^{-1}\rm M_{\odot}$, $r$ is the central-to-total stellar mass ratio, $\log M_\mathrm{*,MMG}/M_\mathrm{*,tot}$, of descendant halos. The $M_\mathrm{*,tot}$ is calculated by the sum of stellar mass for all member galaxies. In \citet{Wang2023}, the derived $\log R_c$ from $m$ and $r$ is consistent with the true $\log R_c$ well with an uncertainty smaller than 0.1 dex.

\begin{figure}
    \centering
    \includegraphics[width=0.48\textwidth]{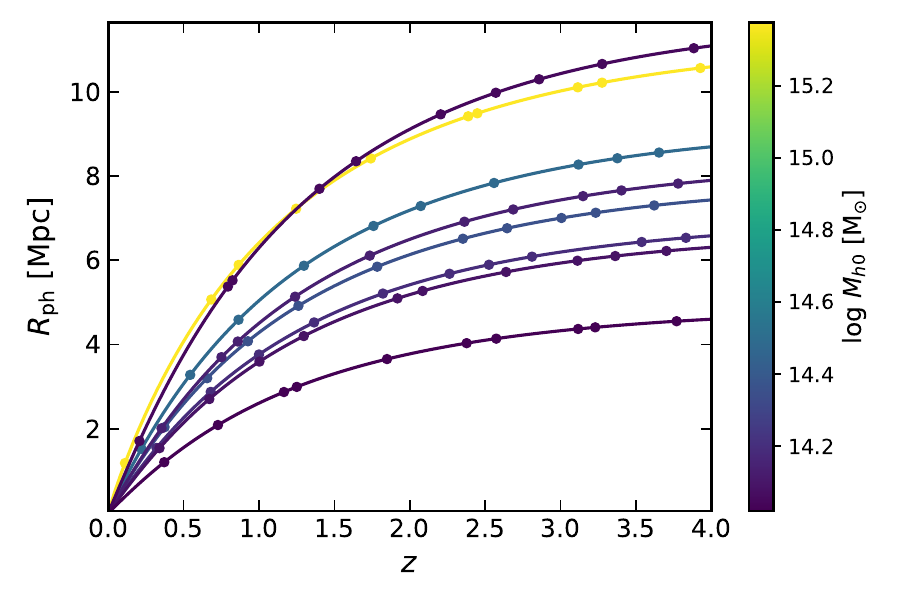}
    \caption{The evolution of protohalo size for the 8 referece clusters as a function of redshift. Each line represents the evolution of galaxies for a progenitor group, with color coded by the halo mass of reference clusters. The points represent the redshift position of the matched progenitors of groups at different redshift bins.}
    \label{fig:Rph}
\end{figure}

In our work, we regard the reference clusters as the descendant halos. We show the protohalo size history of the reference clusters in Figure~\ref{fig:Rph}. In general, more massive clusters have a larger amplitude of protohalo size compared with less massive ones. The growth slope of the protohalo size becomes shallow since $z > 2$ as expected.

\begin{figure*}
    \centering
    \includegraphics[angle=270, width=0.98\textwidth]{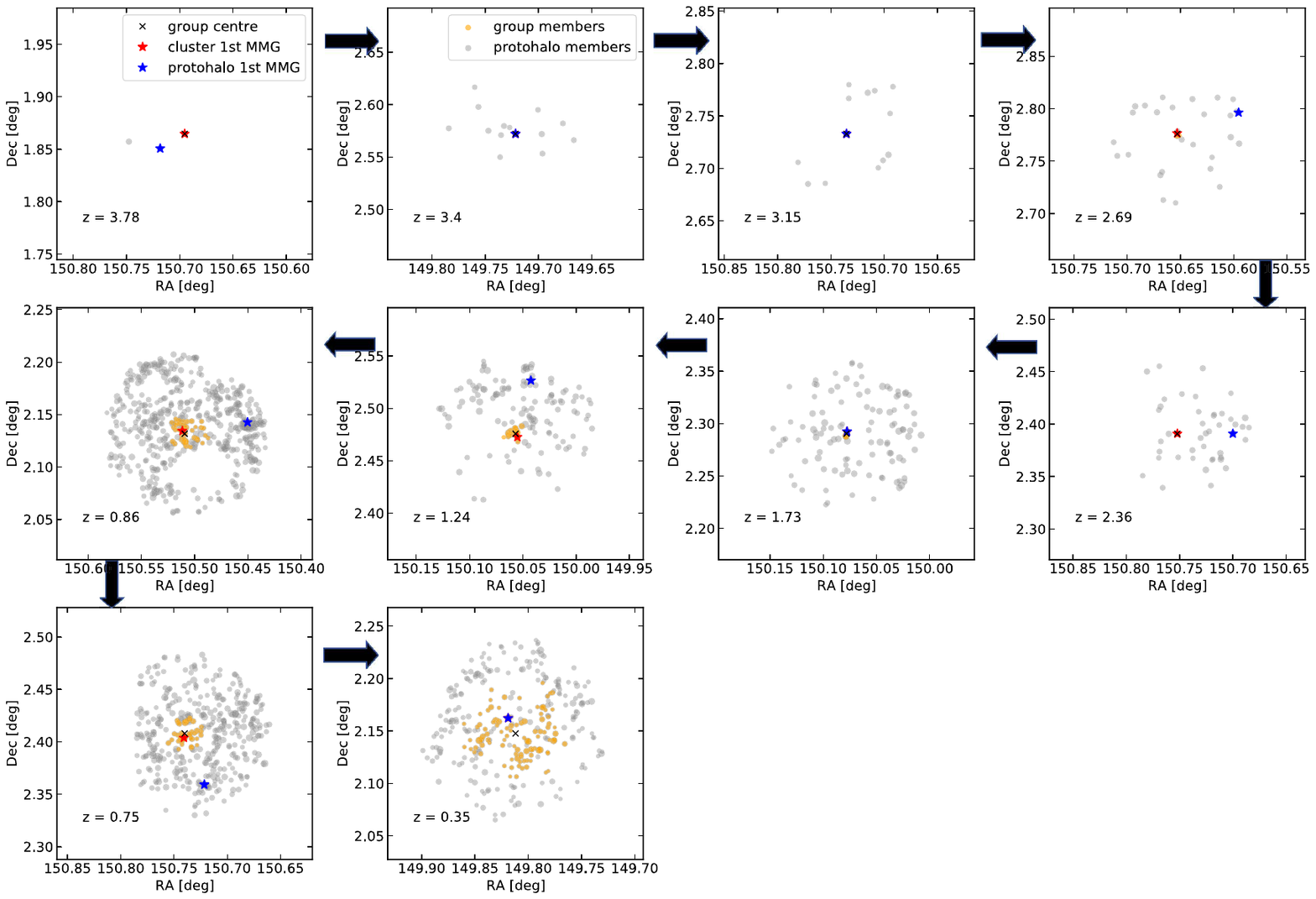}
    \caption{One example of member galaxies distribution in the evolution sequence of clusters and protohalos from $z = 3.78$ to $z = 0.35$. This reference cluster (ID 7) at $z = 0.35$ has a halo mass of $10^{14.36}\ \rm M_{\odot}$. The black arrows mark the evolution trend from high redshift to low redshift. The black cross is the luminosity-weighted group center defined by group finder. The grey and orange points show the member galaxies of groups and protohalos, with the point size proportional to stellar mass. The red and blue stars represent cluster MMGs and protohalo MMGs, respectively.}
    \label{fig:evo_one}
\end{figure*}

Then, we determine the member galaxies of protohalos by taking the progenitors of reference clusters as the cores of protohalos. We used the projected distance perpendicular to the line of sight (\rp) to measure the separation between galaxies and the progenitor of groups. The depth of the selection region is centered at the redshift of the progenitors of groups with a redshift difference of 0.1, which corresponds to the uncertainty of \photoz in the magnitude bin $24 < i \leq 25$. \rp is free from the effect of redshift distortions along the line of sight, defined as (see \citealt{Galaxybook} for a reference):
\begin{equation}
    r_\mathrm{p} = \sqrt{\mathbf{s}^2 - \frac{(\mathbf{s} \cdot \mathbf{l})^2}{|\mathbf{l}|^2}},
\end{equation}
where $\mathbf{s}$ is the separations between protohalos and around groups, which is defined as $\mathbf{s}=\mathbf{s}_1 - \mathbf{s}_2$\footnote{$\mathbf{s}_1$ and $\mathbf{s}_2$ represent the position vector from the Earth to satellite galaxies and group center, respectively.}. $\mathbf{l}$ equals to $(\mathbf{s}_1 + \mathbf{s}_2)/2$. Galaxies or groups with $r_p < R_{\rm ph}$ are identified as part of protohalos. In our analysis, we consider two types of environment in which MMGs survive: the main progenitors of groups and the protohalos. For clarity, we call MMGs in the main progenitors of clusters and protohalos, cluster MMGs and protohalo MMGs, respectively.   

In Figure~\ref{fig:evo_one}, we show an example of the spatial distribution of member galaxies within the progenitors of groups and protohalos of one reference cluster. As the redshift increases, there is a decline in the number of member galaxies within both group progenitors and protohalos. Meanwhile, it is observed that MMGs correspond to identical entities in both progenitor groups and protohalos. 
As we have checked, in most cases, the progenitor groups and protohalos share the same first-order MMGs. However, the second MMGs in protohalos are usually different from those in the progenitor groups.  


\section{Evolution of the MMGs in clusters and protohalos} \label{sec:results}

In this section, we investigate the evolution of the cluster MMGs and protohalo MMGs by analyzing several key physical properties, including stellar mass, luminosity, color, SFR, sSFR, and stellar age, from $z = 4$ to $z = 0$. 
We investigate the evolutionary trends for both the first and second MMGs in clusters and protohalos, with particular attention not only to the evolution of the second MMGs, but also to whether their trends follow a similarity as that of the first MMGs.    

\subsection{Stellar mass}

\begin{figure*}
    \centering
    \includegraphics[width=0.98\textwidth]{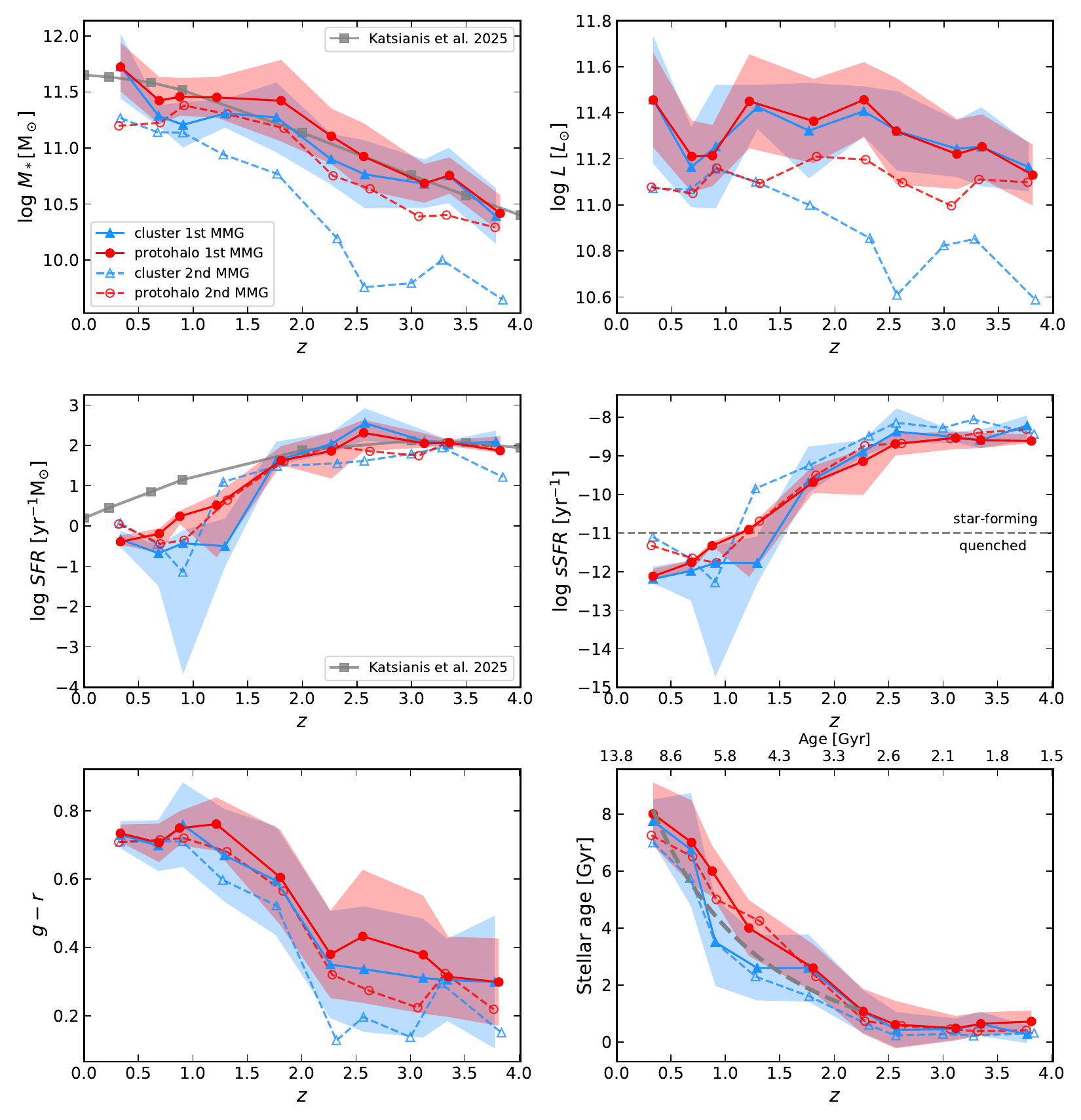}
    \caption{The evolution of physical properties of MMGs, including stellar mass, luminosity, SFR, sSFR, color and stellar age. The blue triangles and red points show the medians for cluster and protohalo MMGs. The solid and hollow symbols represent the physical properties of the first and second MMGs, respectively. The blue and red shaded regions show the $1\sigma$ error for the cluster and protohalo first MMGs, except for SFR and sSFR, where the shaded areas represent the 25th and 75th percentile errors.
    In the top left panel, the grey line with squared markers indicates the theoretical prediction of stellar mass growth for a galaxy with a stellar mass of $10^{11.65}\ \rm M_{\odot}$ at $z = 0$, based on the formula provided by \citet{Katsianis2025}. 
    The luminosity is derived at $i$ band with Equation~\ref{eq:lum}. 
    In the middle left panel, the grey line with squared markers represents a theoretical SFR prediction, derived by converting the stellar mass growth of a galaxy shown in the top left panel using the SFR-$M_*$ relation from \citet{Tomczak2016}. 
    The grey dashed line at $10^{-11}\ \rm yr^{-1}$ in the middle right panel marks the boundary between quenched and star-formation galaxies following the criterion adopted by \citet{Katsianis2019}.
    The color is defined as the rest-frame absolute magnitude difference between $g$ and $r$ bands.
    The upper $x$-axis in the bottom right panel marks the age of the Universe corresponding to redshift, and the dashed gray line represents the increasing age of the Universe on the basis of the median MMGs age at $z = 2.27$.
    }
    \label{fig:mmg_prop}
\end{figure*}

The stellar mass is among the most essential parameters for describing the properties of galaxies. The top left panel of Figure~\ref{fig:mmg_prop} illustrates the progression of stellar mass within both cluster MMGs and protohalo MMGs. We first focus on the first MMGs. As the redshift diminished from roughly $z \sim 4$ to $z \sim 2.5$, the stellar mass of MMGs grew at a nearly steady pace. Subsequently, between $1.5 \lesssim z \lesssim 2.5$, MMGs experienced modest growth. Afterward, the stellar mass starts to ascend sharply again for $z \lesssim 1.0$, peaking at about $\sim 10^{11.75}\ \rm M_{\odot}$ during the final stage. The overall pattern is quite comparable for both cluster and protohalo MMGs. This aligns with the prevailing galaxy formation theories that suggest continued growth in the mass of central galaxies within clusters. However, the growth rate decelerates at around $z \sim 1.8$, marking the peak of galaxy evolution. We attribute this to the combined effects of frequent mergers and intensified galaxy-scale activities like stellar feedback, which impede the usual accretion onto MMGs. Further analysis of the components contributing to the rise in stellar mass can be found in Section~\ref{sec:two_phase}. 

The stellar mass growth of MMGs has been well established in the analytic formula. For example, under the framework of the conditional luminosity function of \citet{Yang2003}, \citet{Yang2012} analytically derived for the first time the mass growth history of central galaxies in dark halos of different masses, combined with the observation of the stellar mass function at different redshifts. In the top left panel of Figure~\ref{fig:mmg_prop}, we include a theoretical prediction of stellar mass growth for a galaxy with a stellar mass of $10^{11.65}\ \rm M_{\odot}$ at $z = 0$, based on the $\Gamma$ shaped evolution of the cosmic star formation rate density from \citet{Katsianis2025}. In general, this theoretical curve follows the observed trend, although it predicts that the MMG forms slightly fewer stars per year at lower redshifts.

In the late time of the Universe ($z \lesssim 1$), there has existed significant disagreement between models and observations \citep[e.g.,][]{Ruszkowski2009, Tonini2012, Lin2013, Shankar2015}. Some studies, both theoretical and observational, suggest that dry mergers (the merging of gas-poor galaxies) contribute to a stellar mass increase of MMGs by a factor of 2$-$3. Through semi-analytic models, \citet{DeLucia2007} predicted that the stellar mass of MMGs would triple from $z = 1$ to the present. In contrast, \citet{Shankar2015} demonstrated using a semi-empirical model that this mass continues to grow up to the current epoch, whereas \citet{Tonini2012} observed a growth plateau starting at $z \sim 0.4$, corroborated by \citet{Bellstedt2016}'s observational study. Our findings indicate a stellar mass increase by a factor of 2.8 from redshift 0.68 to 0.33.

The median evolution of stellar mass for the second MMGs in both clusters and protohalos is depicted in the top left panel of Figure~\ref{fig:mmg_prop} using blue and red lines, respectively. Both cluster-based and protohalo second MMGs exhibit an upward trend in stellar mass from redshift around $\sim3.7$ to 1.5, similar to the observed pattern of the first MMGs. Contrary to the similar mass levels between the cluster and protohalo first MMGs, cluster second MMGs have lower stellar masses than their protohalo counterparts. This discrepancy indicates that at high redshift there are some massive galaxies outside the main halos' virialized zone, which did not end up as the satellite galaxies. The slower growth trend at redshifts below 1.6 may be attributed to a decline in star formation during this period. Notably, the stellar mass of cluster and protohalo second MMGs becomes similar at redshifts below 0.5, likely due to two factors: (1) the protohalo radius at lower redshift approaches the virial radius, reducing the count of massive galaxies beyond the cluster; and (2) a lower star formation rate limits the emergence of new massive satellites in later cosmic times. 

\subsection{Luminosity}

Luminosity is another fundamental property of galaxies and can be conveniently measured from observation. The top right panel of Figure~\ref{fig:mmg_prop} illustrates the evolution of luminosity in cluster and protohalo MMGs. Luminosity is computed using the absolute magnitude as presented in Eq.~\ref{eq:lum}. From $z \sim 4$, MMG luminosity progressively increases as the redshift decreases, attributed to ongoing material accumulation that results in increased luminosity. This pattern aligns with the increase in stellar mass depicted in the top left panel of Figure~\ref{fig:mmg_prop}. The luminosity roughly peaks at around $z \sim 2.3$, followed by a subtle reduction. As we approach $z \sim 0.8$, the luminosity experiences another rise, though \citet{Chu2022} showed that the luminosity of MMGs has no evolution up to redshift 0.7. Nonetheless, due to the significant variability in this period, we avoid making extensive interpretations of this late-time increase. In general, protohalo MMGs follow a luminosity evolution trend that parallels that of cluster MMGs.

In general, the luminosity of the second MMGs is lower than that of first MMGs, for both cluster and protohalo MMGs. Satellite galaxies in clusters are subject to stronger environmental effects compared to those in protohalos, leading to suppressed luminosity. 




\subsection{SFR and sSFR}

SFR and sSFR are essential quantities for characterizing the instantaneous stellar mass build-up and star-formation activities of galaxies. These two quantities describe the connection between gas and stars and are closely linked to galaxy key properties such as color, morphology, and stellar age.

In the middle left panel of Figure~\ref{fig:mmg_prop}, we present the evolution of SFR for cluster and protohalo MMGs. In the redshift range $2.5 < z < 4$, both cluster and protohalo MMGs hold a high and nearly constant SFR of $\gtrsim 10^2\ \rm yr^{-1}M_{\odot}$, indicative of an active star-forming phase. This performance is consistent with the standard framework of galaxy evolution, in which high-redshift galaxies generally exist in a star-forming state. 
Below $z \sim 2.5$, the SFR of the MMGs begins to decline dramatically. During this period, the scatter increases significantly, which is consistent with the findings in \citet{Katsianis2019} who studied the diversity of SFR of galaxies through the scatter of the SFR-$\rm M_*$ relation, suggesting the galaxies undergo diverse evolutionary processes and possibly enhanced environmental effects. At $z \lesssim 0.5$, the SFR converges to approximately 0.5 $\rm yr^{-1}M_{\odot}$ and the reduced scatter indicates that the MMGs enter a relatively homogeneous and quiescent phase. Throughout the full redshift range, both cluster and protohalo MMGs show similar SFR evolution trends. 

Theoretically, the evolution of SFR can be established from the stellar mass growth history. For example, \citet{Behroozi2013} and \citet{Moster2013} employed the stellar-to-halo-mass relation to predict SFR based on a combination of numerical simulations and observations. \citet{Yang2013} derived the star formation history based on the conditional stellar mass function in \citet{Yang2012} and made better constraints using low-redshift observations. In the middle left panel of Figure~\ref{fig:mmg_prop}, we convert the theoretical stellar mass evolution as shown in the top left panel of Figure~\ref{fig:mmg_prop} into a SFR based on the SFR-$M_*$ relation from \citet{Tomczak2016}. The theoretical curve matched well with the observations at $z > 2.5$, but an over-estimation appears at the median and low redshifts.

We include the comparison of SFR for the second MMGs in the cluster and protohalo in the middle left panel of Figure~\ref{fig:mmg_prop}. Overall, the evolution trend is similar regardless of the type of galaxies and environments. We find that the SFR of the second MMGs is slightly lower than that of the first MMGs and presents a slight increasing trend from redshift $z \sim 3.8$ to $z \sim 2.5$. This suggests that the second MMGs are formed in a continuous way, resulting in lower stellar mass compared with the first MMGs as shown in the top left panel of Figure~\ref{fig:mmg_prop}. 

The evolution of sSFR is shown in the middle right panel of Figure~\ref{fig:mmg_prop}. The sSFR follows an evolutionary trend similar to that of the SFR, and the patterns observed for cluster MMGs are analogous to protohalo MMGs. The second MMGs show a higher sSFR compared to the first MMGs at almost all of the redshift range, primarily because of their lower stellar mass. Following \citet{Katsianis2019}, we include a horizontal line at $10^{-11}\ \rm yr^{-1}$ to separate galaxies into quenched (below the line) and star-formation (above the line) galaxies. The MMGs remain passive after $z = 1.5$.  \citet{Bonaventura2017} claimed that the optically and near-IR-selected MMGs at $0 < z < 1.8$ are  referred to as red but not dead. The authors noted that these galaxies are not ``red and dead" as a small ongoing star formation persist. They agreed that the bulk of their stellar population appears to have been in place since $z > 2$, as predicted but that they retain a pulse of star formation activity. Interestingly, the star formation activity of MMGs increases again at $z < 0.5$, even though the galaxy is still considered as quenched. We note that this is not a brief rejuvenation episode, as the elevated activity persists over a long period from $z = 0.5$ to $z = 0.0$.





\subsection{Color}

The color of a galaxy serves as a key indicator of its stellar population, star formation activity, metallicity, dust content, and overall evolutionary history.

In the bottom left panel of Figure~\ref{fig:mmg_prop}, we show the evolution of the color of the cluster and protohalo MMGs. The color is defined as the rest-frame AB absolute magnitude difference between the $g$ and $r$ band (in $2''$ aperture). The absolute magnitude is measured by the \textsc{LePhare}. Initially, there is a flat profile in color from redshift $z \sim 4.0$ to $z \sim2.5$, corresponding to a high SFR as shown in the middle of Figure~\ref{fig:mmg_prop}.
As the redshift decreases from $z \sim 2.5$, the value of $g - r$ gradually increases, indicating that the MMGs are becoming redder. Then, the trend of $g - r$ becomes flat since $z \lesssim 1.0$, consistent with the behavior of SFR as shown in the middle left panel of Figure~\ref{fig:mmg_prop}. This indicates that the galaxies at this stage are transitioning to a red and quenched phase, in agreement with the findings from the studies of nearby MMGs. 
The difference in $g - r$ color between cluster and protohalo MMGs remains small throughout the redshift range. 

The investigation of the second MMG shows a trend similar to that of the first MMG. Although second MMGs appear slightly bluer than first MMGs at $z < 3.0$, their SFR is marginally lower, which contrasts with the typical expectation that more actively star-forming galaxies are bluer.


\subsection{Stellar age}

In the bottom right panel of Figure~\ref{fig:mmg_prop}, we present the evolution of the stellar ages of MMGs. The stellar age is derived from the best-fit stellar template in years. At $2.3 \lesssim z < 4$, MMGs show an almost constant stellar age distribution with a value below 1 $\rm Gyr$. This indicates that the stellar populations in MMGs are young and actively forming at this stage, which is consistent with the picture of SFR (the middle left panel of Figure~\ref{fig:mmg_prop}). As the redshift reduces from $z \sim 2$, the stellar ages of MMGs increase alongside the age of the Universe, as shown by the dashed gray line, indicating a reduced rate of new star formation. Both cluster and protohalo MMGs show a similar trend, consistent with the well-established relation between stellar mass and age, i.e., massive galaxies typically host old stellar populations. 

Furthermore, the age evolution of the second MMGs is consistent with that of the first MMGs, although the second MMGs tend to have a slightly younger stellar population at $z < 2.0$. This is in line with the observed color difference, where the second MMGs appear bluer at this period. This trend is more evident in cluster MMGs than in protohalo MMGs.    


\section{Two phase of stellar mass growth} \label{sec:two_phase}

In this section, we investigate the mechanisms driving the growth of stellar mass for MMGs. Using cosmological simulations of galaxy formation, \citet{Oser2010} found that most stellar particles in massive galaxies are formed at high redshift, either deep within the virial radius (3 kpc) near the forming galaxy center, or in small systems outside the virial radius of the galaxy. In addition, \citet{Ragone2018} presented that half of the star particles that end up in the inner 50 kpc for a galaxy formed at $z \sim 3.7$ (about 12 billion years ago), while half of the total stellar mass is produced at $z \sim 1.5$ (about 9 billion years ago). Based on this picture, the growth of stellar mass can be characterized by two phases: in situ and ex situ. The in-situ mass is caused by the star formation that the existing gas inside the galaxy cools and collapses into stars, which can be described by the SFR of galaxies. The ex-situ mass mainly includes the infalling stars that are initially formed in other galaxies through mergers or accretion. Given our observational data on the stellar mass and SFR of MMGs across various redshifts, it becomes relatively straightforward to explore both in-situ and ex-situ stellar mass growth throughout their histories. 

\subsection{First MMG}
\begin{figure*}
    \centering
    \includegraphics[width=0.98\textwidth]{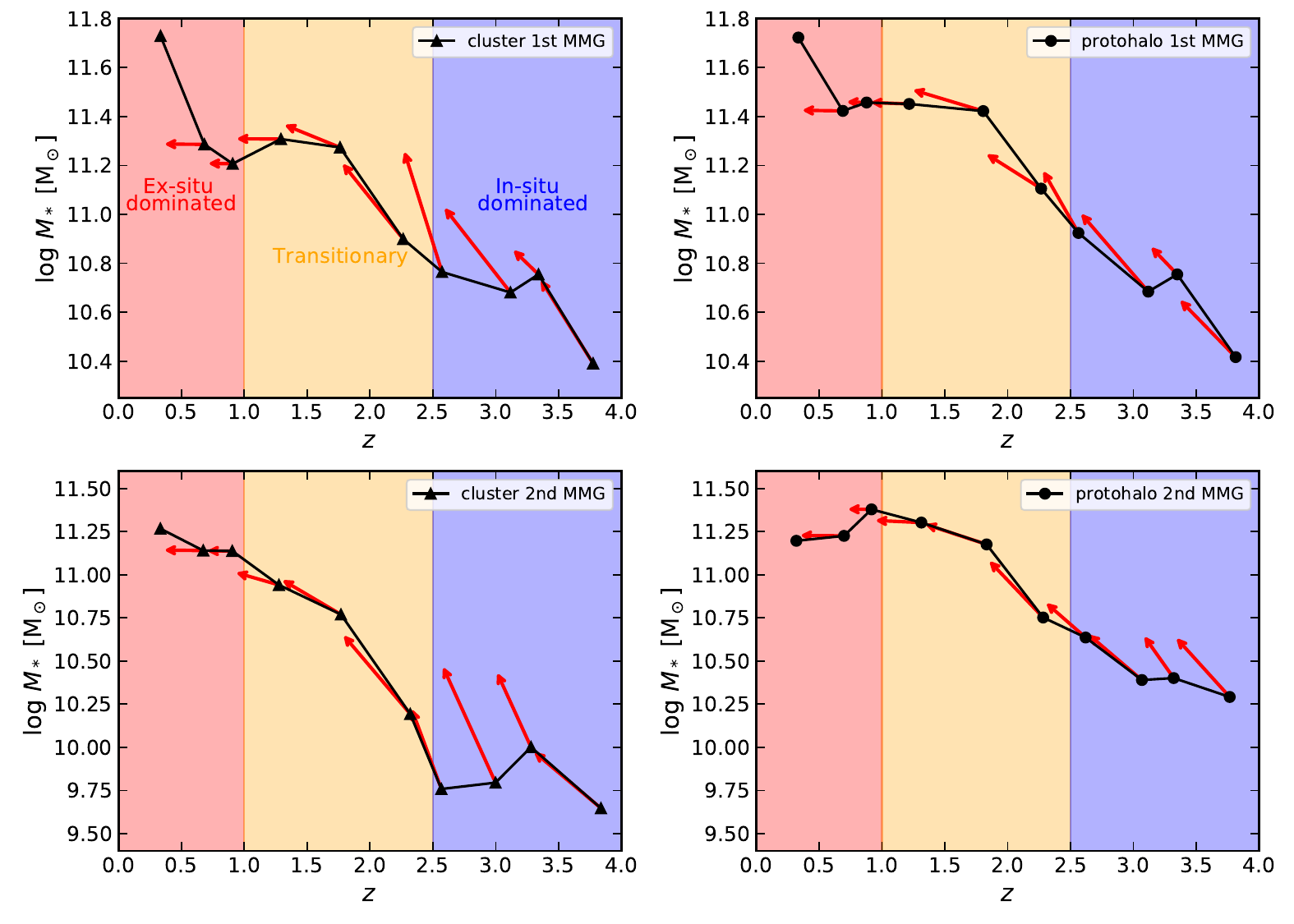}
    \caption{Comparison of stellar mass growth from SED fitting and SFR. The left and right panels show the trends for the cluster and protohalo MMGs, respectively. The top and bottom panels show the comparisons for the first and second MMGs. The black line represents the median stellar mass of the brightest galaxies, while the red arrows indicate the expected stellar mass assuming in-situ growth only, calculated from the SFR and time intervals between redshifts following in Equation~\ref{eq:Msfr}. The shaded region with red, orange, and blue colors indicates three stages of stellar mass growth: in-situ dominated ($z > 2.5$), transitionary ($2.5 \lesssim z \lesssim 1$), and ex-situ dominated ($z < 1$).}
    \label{fig:derive_mass}
\end{figure*}

We illustrate the contributions of in-situ and ex-situ processes to stellar mass growth for cluster and protohalo first MMGs in the top panels of Figure~\ref{fig:derive_mass}. The black line shows the median stellar mass at each redshift, reflecting the combined effect of both in-situ and ex-situ mass growth. The heads of the red arrows indicate the predicted stellar mass, $M_{\rm pi}$, at the next redshift step, derived from the star formation history. Specifically, the predicted stellar mass is calculated using the following equation:
\begin{equation} \label{eq:Msfr}
    M_{\rm pi} = \mathrm{SFR} \times \Delta t + M_{i},
\end{equation}
where $M_{\rm i}$ represents the stellar mass at the current redshift, corresponding to the tails of the red arrows. $\Delta t$ is the time interval between the two redshifts. Therefore, the term $\rm SFR \times \Delta t$ serves as an estimate of the in-situ stellar mass formed during that interval. 

The states of in-situ and ex-situ mass can be divided into three evolutionary stages: 
\begin{itemize}
    \item In-situ dominated time. The first stage, at an early time of $z>2.5$, is that the in-situ mass (indicated by the red arrows) exceeds the actual increase in total stellar mass (shown by the black triangles), suggesting that the ex-situ contribution is effectively negative. This implies that strong feedback processes such as supernovae or AGN-driven winds expel a significant amount of gas or even stars, suppressing overall mass growth. We note that this is not entirely in agreement with what is typically adopted in cosmological simulations. For example, AGN feedback in the EAGLE simulations is constructed to play an important role in shaping star formation at $z < 2$ and is constructed to be relatively minor at $z > 2$ \citep{Katsianis2017}. Another potential reason for the overestimation from Equation~\ref{eq:Msfr} is that the SFRs of galaxies in this phase are not constant, i.e., periods of reduced star formation are interspersed with high star formation episodes, making the assumptions of Equation~\ref{eq:Msfr} invalid.  We note that the COSMOS2020 assumes a fairly smooth star formation history for stellar mass and SFR. Such an assumption would likely lead to underestimated stellar masses and overestimated SFRs if SFRs change on rapid timescales, particularly over this short timespan (1.1 Gyr from redshift 4 to 2.5).

    \item Transitionary epoch. At the second state ($ 1.0 \lesssim z \lesssim 2.5$), the in-situ mass closely matches the observed increase in total stellar mass. This means a balance between star formation and feedback, where stellar mass growth is primarily driven by in-situ processes, and any mass loss from feedback is roughly compensated by ex-situ accretion. The role of ex-situ accretion is supported by \citet{Sawicki2020}, who showed that massive galaxies can experience significant mass growth through major mergers during this epoch, based on a sample of ultra-massive galaxies at $z \sim 1.6$. The above agrees with the findings of \citet{Bonaventura2017}, who inferred via stacked infrared SEDs that a potentially significant fraction of the MMG stellar mass growth since $z > 2$ can be attributed to star formation processes. They estimated that star formation contributes approximately 40.8 per cent of the average SpARCS MMG stellar mass over this redshift range, with the majority of growth occurring at $z > 1$. In contrast, some semi-analytic models have made differing predictions. For example, \citet{Tonini2012} suggested star-forming MMGs are common at later stages, whereas \citet{DeLucia2007} presented that MMGs passively evolve only through minor dry mergers since $z = 2$. It is interesting that our result is not also in agreement with the predictions of the hydrodynamical simulations presented in \citet{Ragone2018}. According to \citet{Ragone2018}, the contribution of the in-situ SF is minor compared to the global SF occurring at $z > 2$ in all the progenitors that end up into the local MMGs. In this high-redshift regime, other progenitors may compete with or even surpass the main one in mass.

    \item Ex-situ dominated time. At the final stage $z < 1.0$, galaxies continue to grow in stellar mass, but the in-situ contribution remains almost flat, indicating that star formation has largely ceased, i.e., the galaxies are quenched. The continued increase in stellar mass is thus predominantly driven by ex-situ processes, such as accretions or mergers. The picture is also supported by \citet{Williams2024}, who analyzed the growth of the outer regions of galaxies at $0.2 \leq z \leq 1.1$ from CLAUDS and HSC-SSP data. They suggested that the additional stellar halo growth is contributed by the accretion via minor mergers, with this effect being stronger in more massive galaxies. The steep slope of mass increase reflects efficient mass assembly with relatively weak feedback. In addition, MMGs in protohalos show a similar trend as those in clusters, but with a steep slope at the early time.
\end{itemize}


Our results are consistent with the stellar mass growth picture described in \citet{Cooke2019}, who investigated the evolution of MMG progenitors in COSMOS since $z \sim 3$. They identified three evolutionary phases: an early in-situ star formation (SF)-dominated phase at $2.25 < z < 3$, which our findings suggest may extend to $z \sim 4$; a transitionary phase lasting until $z \sim 1.25$ where mass growth is derived by in situ SF and additional stellar mass generation such as mergers (both gas rich and poor); and a low-redshift phase dominated by ex-situ stellar mass growth, particularly through dry mergers. Our results, showing that the stellar mass increase due to in-situ SF is slightly lower than the total mass growth during the transitional epoch, further support this evolutionary picture.


\subsection{Second MMG}

The bottom panels of Figure~\ref{fig:derive_mass} illustrate the comparative growth in stellar mass from both in-situ and ex-situ sources for the second MMGs. Similar to the first MMGs, both cluster and protohalo second MMGs undergo an in-situ star formation driven phase at redshift $z \gtrsim 2.5$ followed by a similar transitionary epoch. However, the second MMGs in clusters differ from the first MMGs as they exhibit prominent stellar mass decrease near $z \sim 2.8$. We hypothesize that this might be caused by the merger of the second MMGs to the first MMGs and intense hydrodynamic phenomena, such as ram pressure stripping and starvation, combined with other environmental influences, impede star formation and overall stellar mass accumulation.  By redshift $z \le 1.25$, both the total and in-situ stellar mass growth of the second most massive galaxies (second MMGs) in clusters and protohalos reach a plateau, indicating a decline in contributions from both in-situ star formation and ex-situ mergers. This suggests that, at lower redshifts, the second MMGs experience fewer mergers and are less favored for satellite accretion compared to the continued growth seen in the most massive galaxies (first MMGs). In addition to these, we note that some massive galaxies in the protohalos can also be accreted into the clusters acting as the new cluster second MMGs. Consequently, along the evolution trajectory, the second MMGs in the clusters are not necessarily the direct progenitors in descendant clusters. 

\section{Discussion} \label{sec:disc}

\begin{figure}
    \centering
    \includegraphics[width=0.48\textwidth]{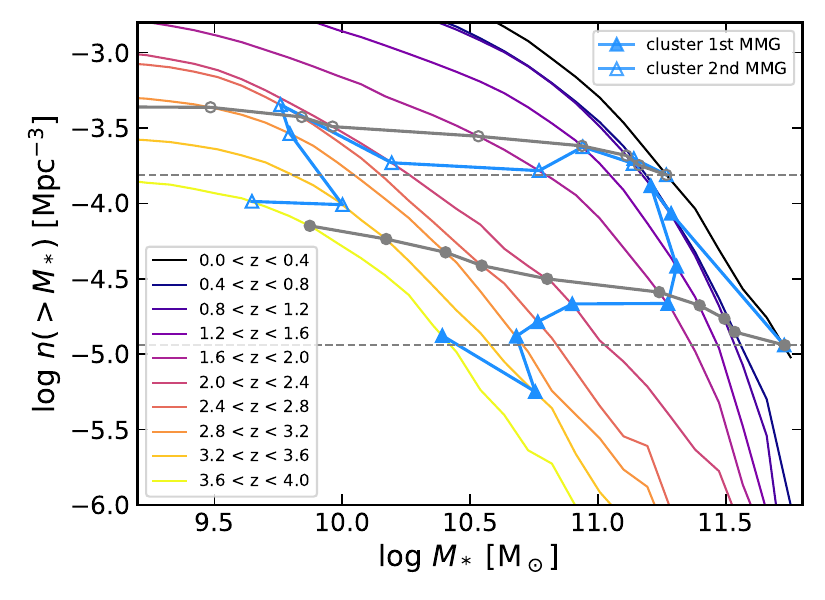}
    \caption{Cumulative stellar mass functions at different redshifts, coded with different color lines. The solid and hollow triangles show the evolution trajectory of cluster first and second galaxies based on the median stellar mass shown in the top left panel of Figure~\ref{fig:mmg_prop}. The solid and hollow gray points indicate the evolution trajectory of the cluster first and second MMGs from the evolving cumulative number density method \citep{Behroozi2013} based on the number density of MMGs at $0 < z < 0.4$. The gray dashed lines mark the level of constant cumulative number density.}
    \label{fig:smf_evo}
\end{figure}

Apart from tracing the ancestors of massive galaxies using clusters/groups, several approaches have been developed to directly follow their progenitors using only galaxies. These methods include studying the evolution of scaling relations \citep{Franx2008}, analyzing the luminosity function \citep{Faber2007}, and examining the constant or changing number density of galaxies \citep{vanDokkum2010, Leja2013, Behroozi2013, Clauwens2016, Hill2017, Cooke2019, Wang2023b}. In this section, we offer a comparison with the evolving number density method, which is frequently used in galaxy evolution studies.

In Figure~\ref{fig:smf_evo}, we present the cumulative galaxy stellar mass functions at different redshifts, together with the evolutionary trajectory of cluster MMGs based on the stellar mass evolution shown in the top left panel of Figure~\ref{fig:mmg_prop} using blue symbols. 
The theoretical evolutionary trajectory of the cluster first and second MMGs is derived from the evolving cumulative number density method \citep{Behroozi2013} based on the number density of MMGs at $0 < z < 0.4$ using gray symbols.
The two methods of matching progenitor galaxies present different trajectories. The trajectory using the evolving cumulative number density method is similar to that presented in \citet{Hill2017} (Figure 1), showing an increasing trend with
redshift overall as expected. The trajectories based on the cluster abundance matching method show an oscillation in the cumulative stellar mass function at different redshifts, implying a complex evolution history of MMGs. This differs from the evolution of general galaxies in the Universe that the number of galaxies with equivalent mass (i.e., number density) decreases with redshift. We propose that this variation might result from MMGs within the clusters being sporadically affected by mergers, significant feedbacks, and hydrodynamical processes, leading to even temporary mass reduction. Consequently, the stellar masses of the progenitors at the highest redshift are significantly higher than those predicted by the evolving cumulative number density method.

\section{Summary} \label{sec:summary}

We investigate the evolution of the bright galaxies, specifically the first brightest and second brightest galaxy, within two cosmic environments, clusters and protohalos from $z = 4$ to $z = 0$ using the COSMOS2020 galaxy catalog. A range of physical properties are explored, including stellar mass, luminosity, color, SFR, sSFR, and stellar age. We begin by identifying galaxy clusters from a selected galaxy sample using the halo-based group finder. The progenitors of the reference clusters ($z < 0.4$) are determined using the abundance matching method, thus constructing the evolution tracks of clusters. Corresponding protohalos are defined by encompassing galaxies within a specified protohalo radius along the evolution chain. In addition, we investigate the stellar mass buildup in cluster and protohalo MMGs, distinguishing between in-situ and ex-situ contributions. We discuss the methods used to trace the evolution of massive galaxies. Our conclusions are summarized as follows:

\begin{itemize}
    \item Throughout the redshift range, the cluster and protohalo first MMGs are always the same objects and follow similar evolutionary trends, while the differences appear between the cluster and protohalo second MMGs in terms of stellar mass, likely due to the environmental effects.  

    \item As the redshift decreases, the stellar mass of MMGs tends to increase, continuing into the nearby Universe. Conversely, second MMGs experience a slower mass growth around $z \lesssim 1.5$, probably because of diminished star formation and the lack of accreting satellite galaxies.

    \item At $ 2.5 \lesssim  z < 4$, MMGs form stars at a nearly constant level of SFR $\sim 10^{2}\ \rm yr^{-1}M_{\odot}$, while their SFR begins to decline since $z \lesssim 2$, showing a sharp decreasing trend of SFR indicative of quenching in the later evolutionary stages. The large uncertainty at the intermediate redshift range suggests that the galaxies undergo diverse evolutionary processes and possibly enhanced environmental effects. Meanwhile, the evolution of sSFR presents a similar trend as SFR. 
    
    \item The color ($g - r$) of MMGs becomes progressively redder from $z \sim 2.5$ and remains consistently red since $z < 1$. The sSFR of MMGs remain always under $10^{-11}\ \rm yr^{-1}$ implying a quenched/passive state.

    \item The stellar population of MMGs is primarily formed at $z > 2$, after which the stellar age increases in parallel with the age of the Universe, indicating a declining rate of newly formed stars at later times.


    \item We identify three stages of the stellar mass growth by dividing the contributing components into two phases: in situ and ex situ. The first stage, at $z > 2.5$, is mainly driven by the in-situ mass, although the feedback effects may reduce the net mass growth. In the second state ($1.0 \lesssim z \lesssim 2.5$), the mass increase contributed by in-situ formation is consistent with the overall stellar mass growth. At the final stage, the mass increase is predominantly driven by ex-situ processes.   

    \item The evolutionary trajectories of cluster MMGs show an oscillation in cumulative stellar mass functions, which differ from the increasing trend with redshift derived from the evolving number density method. The oscillated trajectories imply a more complex evolution history of cluster MMGs.
\end{itemize}

Our work extends the observational study of MMG evolution to $z \sim 4$, incorporating the environment of protohalos and a wide investigation of MMG properties. In addition, we introduce a novel method to construct the evolutionary chains of galaxy clusters in observation, offering new potential for future studies of galaxy evolution. Although our analysis is based on a limited sky area, ongoing and upcoming deep-field surveys such as \emph{Euclid} \citep{Laureijs2011}, \emph{CSST} \citep{Zhan2011,Gong2019} and LSST \citep{Ivezic2019} are expected to yield larger samples of clusters and protoclusters. Furthermore, ongoing large spectroscopic surveys \citep[e.g., PFS, ][]{Takada2014} will enable more accurate measurements of galaxy properties, particularly enhancing cluster identification and the characterization of MMGs. These advancements will allow for more valid examination of MMG evolutionary scenarios. The cluster and protohalo catalogs developed in this work are publicly available online{\footnote{https://gax.sjtu.edu.cn/data/PFS.html} or on Zenodo{\footnote{https://doi.org/10.5281/zenodo.17060220}}.


\section*{Acknowledgements}

This work is supported by the National Key R\&D Program of China (2023YFA1607800, 2023YFA1607804, 2022YFA1605300), the China Manned Space Project with Nos. CMS-CSST-2021-A02 \& CMS-CSST-2025-A04, “the Fundamental Research Funds for the Central Universities”, 111 project No. B20019, and Shanghai Natural Science Foundation, grant No. 19ZR1466800, and the National Nature Science Foundation of China (NSFC) grants No. 12273051. This work has made use of the Gravity Supercomputer at the Department of Astronomy, Shanghai Jiao Tong University. Ilaria Marini and Paola Popesso acknowledge funding from the European Research Council (ERC) under the European Union’s Horizon Europe research and innovation programme ERC CoG (Grant agreement No. 101045437).

\bibliography{paper}{}

\begin{thebibliography}{}
\expandafter\ifx\csname natexlab\endcsname\relax\def\natexlab#1{#1}\fi
\providecommand{\url}[1]{\href{#1}{#1}}
\providecommand{\dodoi}[1]{doi:~\href{http://doi.org/#1}{\nolinkurl{#1}}}
\providecommand{\doeprint}[1]{\href{http://ascl.net/#1}{\nolinkurl{http://ascl.net/#1}}}
\providecommand{\doarXiv}[1]{\href{https://arxiv.org/abs/#1}{\nolinkurl{https://arxiv.org/abs/#1}}}

\bibitem[{{Aihara} {et~al.}(2019){Aihara}, {AlSayyad}, {Ando}, {Armstrong}, {Bosch}, {Egami}, {Furusawa}, {Furusawa}, {Goulding}, {Harikane}, {Hikage}, {Ho}, {Hsieh}, {Huang}, {Ikeda}, {Imanishi}, {Ito}, {Iwata}, {Jaelani}, {Kakuma}, {Kawana}, {Kikuta}, {Kobayashi}, {Koike}, {Komiyama}, {Li}, {Liang}, {Lin}, {Luo}, {Lupton}, {Lust}, {MacArthur}, {Matsuoka}, {Mineo}, {Miyatake}, {Miyazaki}, {More}, {Murata}, {Namiki}, {Nishizawa}, {Oguri}, {Okabe}, {Okamoto}, {Okura}, {Ono}, {Onodera}, {Onoue}, {Osato}, {Ouchi}, {Shibuya}, {Strauss}, {Sugiyama}, {Suto}, {Takada}, {Takagi}, {Takata}, {Takita}, {Tanaka}, {Terai}, {Toba}, {Uchiyama}, {Utsumi}, {Wang}, {Wang}, \& {Yamada}}]{Aihara2019}
{Aihara}, H., {AlSayyad}, Y., {Ando}, M., {et~al.} 2019, \pasj, 71, 114, \dodoi{10.1093/pasj/psz103}

\bibitem[{{Ando} {et~al.}(2022){Ando}, {Shimasaku}, {Momose}, {Ito}, {Sawicki}, \& {Shimakawa}}]{Ando2022}
{Ando}, M., {Shimasaku}, K., {Momose}, R., {et~al.} 2022, \mnras, \dodoi{10.1093/mnras/stac1049}

\bibitem[{{Arnouts} {et~al.}(2002){Arnouts}, {Moscardini}, {Vanzella}, {Colombi}, {Cristiani}, {Fontana}, {Giallongo}, {Matarrese}, \& {Saracco}}]{Arnouts2002}
{Arnouts}, S., {Moscardini}, L., {Vanzella}, E., {et~al.} 2002, \mnras, 329, 355, \dodoi{10.1046/j.1365-8711.2002.04988.x}

\bibitem[{{Baraffe} {et~al.}(2015){Baraffe}, {Homeier}, {Allard}, \& {Chabrier}}]{Baraffe2015}
{Baraffe}, I., {Homeier}, D., {Allard}, F., \& {Chabrier}, G. 2015, \aap, 577, A42, \dodoi{10.1051/0004-6361/201425481}

\bibitem[{{Behroozi} {et~al.}(2013){Behroozi}, {Wechsler}, \& {Conroy}}]{Behroozi2013}
{Behroozi}, P.~S., {Wechsler}, R.~H., \& {Conroy}, C. 2013, \apj, 770, 57, \dodoi{10.1088/0004-637X/770/1/57}

\bibitem[{{Bellstedt} {et~al.}(2016){Bellstedt}, {Lidman}, {Muzzin}, {Franx}, {Guatelli}, {Hill}, {Hoekstra}, {Kurinsky}, {Labbe}, {Marchesini}, {Marsan}, {Safavi-Naeini}, {Sif{\'o}n}, {Stefanon}, {van de Sande}, {van Dokkum}, \& {Weigel}}]{Bellstedt2016}
{Bellstedt}, S., {Lidman}, C., {Muzzin}, A., {et~al.} 2016, \mnras, 460, 2862, \dodoi{10.1093/mnras/stw1184}

\bibitem[{{Bohlin} {et~al.}(1995){Bohlin}, {Colina}, \& {Finley}}]{Bohlin1995}
{Bohlin}, R.~C., {Colina}, L., \& {Finley}, D.~S. 1995, \aj, 110, 1316, \dodoi{10.1086/117606}

\bibitem[{{Bonaventura} {et~al.}(2017){Bonaventura}, {Webb}, {Muzzin}, {Noble}, {Lidman}, {Wilson}, {Yee}, {Geach}, {Hezaveh}, {Shupe}, \& {Surace}}]{Bonaventura2017}
{Bonaventura}, N.~R., {Webb}, T.~M.~A., {Muzzin}, A., {et~al.} 2017, \mnras, 469, 1259, \dodoi{10.1093/mnras/stx722}

\bibitem[{{Brammer} {et~al.}(2008){Brammer}, {van Dokkum}, \& {Coppi}}]{Brammer2008}
{Brammer}, G.~B., {van Dokkum}, P.~G., \& {Coppi}, P. 2008, \apj, 686, 1503, \dodoi{10.1086/591786}

\bibitem[{{Bruzual} \& {Charlot}(2003)}]{Bruzual2003}
{Bruzual}, G., \& {Charlot}, S. 2003, \mnras, 344, 1000, \dodoi{10.1046/j.1365-8711.2003.06897.x}

\bibitem[{{Cai} {et~al.}(2019){Cai}, {Cantalupo}, {Prochaska}, {Arrigoni Battaia}, {Burchett}, {Li}, {Chisholm}, {Bundy}, \& {Hennawi}}]{Cai2019}
{Cai}, Z., {Cantalupo}, S., {Prochaska}, J.~X., {et~al.} 2019, \apjs, 245, 23, \dodoi{10.3847/1538-4365/ab4796}

\bibitem[{{Cerulo} {et~al.}(2019){Cerulo}, {Orellana}, \& {Covone}}]{Cerulo2019}
{Cerulo}, P., {Orellana}, G.~A., \& {Covone}, G. 2019, \mnras, 487, 3759, \dodoi{10.1093/mnras/stz1495}

\bibitem[{Chabrier {et~al.}(2000)Chabrier, Baraffe, Allard, \& Hauschildt}]{Chabrier2000}
Chabrier, G., Baraffe, I., Allard, F., \& Hauschildt, P. 2000, The Astrophysical Journal, 542, 464, \dodoi{10.1086/309513}

\bibitem[{{Chiang} {et~al.}(2013){Chiang}, {Overzier}, \& {Gebhardt}}]{Chiang2013}
{Chiang}, Y.-K., {Overzier}, R., \& {Gebhardt}, K. 2013, \apj, 779, 127, \dodoi{10.1088/0004-637X/779/2/127}

\bibitem[{{Chu} {et~al.}(2021){Chu}, {Durret}, \& {M{\'a}rquez}}]{Chu2021}
{Chu}, A., {Durret}, F., \& {M{\'a}rquez}, I. 2021, \aap, 649, A42, \dodoi{10.1051/0004-6361/202040245}

\bibitem[{{Chu} {et~al.}(2022){Chu}, {Sarron}, {Durret}, \& {M{\'a}rquez}}]{Chu2022}
{Chu}, A., {Sarron}, F., {Durret}, F., \& {M{\'a}rquez}, I. 2022, \aap, 666, A54, \dodoi{10.1051/0004-6361/202243504}

\bibitem[{{Clauwens} {et~al.}(2016){Clauwens}, {Franx}, \& {Schaye}}]{Clauwens2016}
{Clauwens}, B., {Franx}, M., \& {Schaye}, J. 2016, \mnras, 463, L1, \dodoi{10.1093/mnrasl/slw137}

\bibitem[{{Collins} \& {Mann}(1998)}]{Collins1998}
{Collins}, C.~A., \& {Mann}, R.~G. 1998, \mnras, 297, 128, \dodoi{10.1046/j.1365-8711.1998.01482.x}

\bibitem[{{Cooke} {et~al.}(2018){Cooke}, {Fogarty}, {Kartaltepe}, {Moustakas}, {O'Dea}, \& {Postman}}]{Cooke2018}
{Cooke}, K.~C., {Fogarty}, K., {Kartaltepe}, J.~S., {et~al.} 2018, \apj, 857, 122, \dodoi{10.3847/1538-4357/aab895}

\bibitem[{{Cooke} {et~al.}(2019){Cooke}, {Kartaltepe}, {Tyler}, {Darvish}, {Casey}, {Le F{\`e}vre}, {Salvato}, \& {Scoville}}]{Cooke2019}
{Cooke}, K.~C., {Kartaltepe}, J.~S., {Tyler}, K.~D., {et~al.} 2019, \apj, 881, 150, \dodoi{10.3847/1538-4357/ab30c9}

\bibitem[{{De Lucia} \& {Blaizot}(2007)}]{DeLucia2007}
{De Lucia}, G., \& {Blaizot}, J. 2007, \mnras, 375, 2, \dodoi{10.1111/j.1365-2966.2006.11287.x}

\bibitem[{Edward {et~al.}(2023)Edward, Balogh, Bahe, Cooper, Hatch, Marchioni, Muzzin, Noble, Rednick, Vulcani, Wilson, Lucia, Demarco, Forrest, Hirschmann, Castignani, Cerulo, Finn, Hewitt, Jablonka, Kodama, Maurogordato, Nantais, \& Xie}]{Edward2023}
Edward, A.~H., Balogh, M.~L., Bahe, Y.~M., {et~al.} 2023

\bibitem[{{Euclid Collaboration} {et~al.}(2022){Euclid Collaboration}, {Moneti}, {McCracken}, {Shuntov}, {Kauffmann}, {Capak}, {Davidzon}, {Ilbert}, {Scarlata}, {Toft}, {Weaver}, {Chary}, {Cuby}, {Faisst}, {Masters}, {McPartland}, {Mobasher}, {Sanders}, {Scaramella}, {Stern}, {Szapudi}, {Teplitz}, {Zalesky}, {Amara}, {Auricchio}, {Bodendorf}, {Bonino}, {Branchini}, {Brau-Nogue}, {Brescia}, {Brinchmann}, {Capobianco}, {Carbone}, {Carretero}, {Castander}, {Castellano}, {Cavuoti}, {Cimatti}, {Cledassou}, {Congedo}, {Conselice}, {Conversi}, {Copin}, {Corcione}, {Costille}, {Cropper}, {Da Silva}, {Degaudenzi}, {Douspis}, {Dubath}, {Duncan}, {Dupac}, {Dusini}, {Farrens}, {Ferriol}, {Fosalba}, {Frailis}, {Franceschi}, {Fumana}, {Garilli}, {Gillis}, {Giocoli}, {Granett}, {Grazian}, {Grupp}, {Haugan}, {Hoekstra}, {Holmes}, {Hormuth}, {Hudelot}, {Jahnke}, {Kermiche}, {Kiessling}, {Kilbinger}, {Kitching}, {Kohley}, {K{\"u}mmel}, {Kunz}, {Kurki-Suonio}, {Ligori}, {Lilje}, {Lloro}, {Maiorano}, {Mansutti}, {Marggraf},
  {Markovic}, {Marulli}, {Massey}, {Maurogordato}, {Meneghetti}, {Merlin}, {Meylan}, {Moresco}, {Moscardini}, {Munari}, {Niemi}, {Padilla}, {Paltani}, {Pasian}, {Pedersen}, {Pires}, {Poncet}, {Popa}, {Pozzetti}, {Raison}, {Rebolo}, {Rhodes}, {Rix}, {Roncarelli}, {Rossetti}, {Saglia}, {Schneider}, {Secroun}, {Seidel}, {Serrano}, {Sirignano}, {Sirri}, {Stanco}, {Tallada-Cresp{\'\i}}, {Taylor}, {Tereno}, {Toledo-Moreo}, {Torradeflot}, {Wang}, {Welikala}, {Weller}, {Zamorani}, {Zoubian}, {Andreon}, {Bardelli}, {Camera}, {Graci{\'a}-Carpio}, {Medinaceli}, {Mei}, {Polenta}, {Romelli}, {Sureau}, {Tenti}, {Vassallo}, {Zacchei}, {Zucca}, {Baccigalupi}, {Balaguera-Antol{\'\i}nez}, {Bernardeau}, {Biviano}, {Bolzonella}, {Bozzo}, {Burigana}, {Cabanac}, {Cappi}, {Carvalho}, {Casas}, {Castignani}, {Colodro-Conde}, {Coupon}, {Courtois}, {Di Ferdinando}, {Farina}, {Finelli}, {Flose-Reimberg}, {Fotopoulou}, {Galeotta}, {Ganga}, {Garcia-Bellido}, {Gaztanaga}, {Gozaliasl}, {Hook}, {Joachimi}, {Kansal}, {Keihanen},
  {Kirkpatrick}, {Lindholm}, {Mainetti}, {Maino}, {Maoli}, {Martinelli}, {Martinet}, {Maturi}, {Metcalf}, {Morgante}, {Morisset}, {Nucita}, {Patrizii}, {Potter}, {Renzi}, {Riccio}, {S{\'a}nchez}, {Sapone}, {Schirmer}, {Schultheis}, {Scottez}, {Sefusatti}, {Teyssier}, {Tubio}, {Tutusaus}, {Valiviita}, {Viel}, \& {Hildebrandt}}]{Moneti2022}
{Euclid Collaboration}, {Moneti}, A., {McCracken}, H.~J., {et~al.} 2022, \aap, 658, A126, \dodoi{10.1051/0004-6361/202142361}

\bibitem[{{Faber} {et~al.}(2007){Faber}, {Willmer}, {Wolf}, {Koo}, {Weiner}, {Newman}, {Im}, {Coil}, {Conroy}, {Cooper}, {Davis}, {Finkbeiner}, {Gerke}, {Gebhardt}, {Groth}, {Guhathakurta}, {Harker}, {Kaiser}, {Kassin}, {Kleinheinrich}, {Konidaris}, {Kron}, {Lin}, {Luppino}, {Madgwick}, {Meisenheimer}, {Noeske}, {Phillips}, {Sarajedini}, {Schiavon}, {Simard}, {Szalay}, {Vogt}, \& {Yan}}]{Faber2007}
{Faber}, S.~M., {Willmer}, C.~N.~A., {Wolf}, C., {et~al.} 2007, \apj, 665, 265, \dodoi{10.1086/519294}

\bibitem[{{Franx} {et~al.}(2008){Franx}, {van Dokkum}, {F{\"o}rster Schreiber}, {Wuyts}, {Labb{\'e}}, \& {Toft}}]{Franx2008}
{Franx}, M., {van Dokkum}, P.~G., {F{\"o}rster Schreiber}, N.~M., {et~al.} 2008, \apj, 688, 770, \dodoi{10.1086/592431}

\bibitem[{{Fraser-McKelvie} {et~al.}(2014){Fraser-McKelvie}, {Brown}, \& {Pimbblet}}]{Fraser2014}
{Fraser-McKelvie}, A., {Brown}, M.~J.~I., \& {Pimbblet}, K.~A. 2014, \mnras, 444, L63, \dodoi{10.1093/mnrasl/slu117}

\bibitem[{{Fukushima} {et~al.}(2023){Fukushima}, {Nagamine}, \& {Shimizu}}]{Fukushima2023}
{Fukushima}, K., {Nagamine}, K., \& {Shimizu}, I. 2023, \mnras, 525, 3760, \dodoi{10.1093/mnras/stad2526}

\bibitem[{{Gong} {et~al.}(2019){Gong}, {Liu}, {Cao}, {Chen}, {Fan}, {Li}, {Li}, {Li}, {Zhang}, \& {Zhan}}]{Gong2019}
{Gong}, Y., {Liu}, X., {Cao}, Y., {et~al.} 2019, \apj, 883, 203, \dodoi{10.3847/1538-4357/ab391e}

\bibitem[{{Gozaliasl} {et~al.}(2024){Gozaliasl}, {Finoguenov}, {Babul}, {Ilbert}, {Sargent}, {Vardoulaki}, {Faisst}, {Liu}, {Shuntov}, {Cooper}, {Dolag}, {Toft}, {Magdis}, {Toni}, {Mobasher}, {Barr{\'e}}, {Cui}, \& {Rennehan}}]{Gozalias2024}
{Gozaliasl}, G., {Finoguenov}, A., {Babul}, A., {et~al.} 2024, \aap, 690, A315, \dodoi{10.1051/0004-6361/202449543}

\bibitem[{{Han} {et~al.}(2025){Han}, {Li}, {Jiang}, {Chen}, {Wang}, {Wei}, {He}, {He}, {Zhang}, {Liu}, {Cui}, {Gu}, {Guo}, {Jing}, {Kang}, {Li}, {Luo}, {Luo}, {Pei}, {Qiu}, {Tan}, {Xie}, {Yang}, {Yu}, {Yu}, \& {Zhou}}]{Han2025}
{Han}, J., {Li}, M., {Jiang}, W., {et~al.} 2025, arXiv e-prints, arXiv:2503.21368, \dodoi{10.48550/arXiv.2503.21368}

\bibitem[{{Harvey} {et~al.}(2019){Harvey}, {Robertson}, {Massey}, \& {McCarthy}}]{Harvey2019}
{Harvey}, D., {Robertson}, A., {Massey}, R., \& {McCarthy}, I.~G. 2019, \mnras, 488, 1572, \dodoi{10.1093/mnras/stz1816}

\bibitem[{{Hill} {et~al.}(2017){Hill}, {Muzzin}, {Franx}, {Clauwens}, {Schreiber}, {Marchesini}, {Stefanon}, {Labbe}, {Brammer}, {Caputi}, {Fynbo}, {Milvang-Jensen}, {Skelton}, {van Dokkum}, \& {Whitaker}}]{Hill2017}
{Hill}, A.~R., {Muzzin}, A., {Franx}, M., {et~al.} 2017, \apj, 837, 147, \dodoi{10.3847/1538-4357/aa61fe}

\bibitem[{{Hsieh} {et~al.}(2012){Hsieh}, {Wang}, {Hsieh}, {Lin}, {Yan}, {Lim}, \& {Ho}}]{Hsieh2012}
{Hsieh}, B.-C., {Wang}, W.-H., {Hsieh}, C.-C., {et~al.} 2012, \apjs, 203, 23, \dodoi{10.1088/0067-0049/203/2/23}

\bibitem[{{Hu} {et~al.}(2021){Hu}, {Wang}, {Infante}, {Rhoads}, {Zheng}, {Yang}, {Malhotra}, {Barrientos}, {Jiang}, {Gonz{\'a}lez-L{\'o}pez}, {Prieto}, {Perez}, {Hibon}, {Galaz}, {Coughlin}, {Harish}, {Kong}, {Kang}, {Khostovan}, {Pharo}, {Valdes}, {Wold}, {Walker}, \& {Zheng}}]{Hu2021}
{Hu}, W., {Wang}, J., {Infante}, L., {et~al.} 2021, Nature Astronomy, 5, 485, \dodoi{10.1038/s41550-020-01291-y}

\bibitem[{{Ilbert} {et~al.}(2006){Ilbert}, {Arnouts}, {McCracken}, {Bolzonella}, {Bertin}, {Le F{\`e}vre}, {Mellier}, {Zamorani}, {Pell{\`o}}, {Iovino}, {Tresse}, {Le Brun}, {Bottini}, {Garilli}, {Maccagni}, {Picat}, {Scaramella}, {Scodeggio}, {Vettolani}, {Zanichelli}, {Adami}, {Bardelli}, {Cappi}, {Charlot}, {Ciliegi}, {Contini}, {Cucciati}, {Foucaud}, {Franzetti}, {Gavignaud}, {Guzzo}, {Marano}, {Marinoni}, {Mazure}, {Meneux}, {Merighi}, {Paltani}, {Pollo}, {Pozzetti}, {Radovich}, {Zucca}, {Bondi}, {Bongiorno}, {Busarello}, {de La Torre}, {Gregorini}, {Lamareille}, {Mathez}, {Merluzzi}, {Ripepi}, {Rizzo}, \& {Vergani}}]{Ilbert2006}
{Ilbert}, O., {Arnouts}, S., {McCracken}, H.~J., {et~al.} 2006, \aap, 457, 841, \dodoi{10.1051/0004-6361:20065138}

\bibitem[{Ilbert {et~al.}(2008)Ilbert, Capak, Salvato, Aussel, McCracken, Sanders, Scoville, Kartaltepe, Arnouts, Floc'h, Mobasher, Taniguchi, Lamareille, Leauthaud, Sasaki, Thompson, Zamojski, Zamorani, Bardelli, Bolzonella, Bongiorno, Brusa, Caputi, Carollo, Contini, Cook, Coppa, Cucciati, de~la Torre, de~Ravel, Franzetti, Garilli, Hasinger, Iovino, Kampczyk, Kneib, Knobel, Kovac, Borgne, Brun, Fèvre, Lilly, Looper, Maier, Mainieri, Mellier, Mignoli, Murayama, Pellò, Peng, Pérez-Montero, Renzini, Ricciardelli, Schiminovich, Scodeggio, Shioya, Silverman, Surace, Tanaka, Tasca, Tresse, Vergani, \& Zucca}]{Ilbert2009}
Ilbert, O., Capak, P., Salvato, M., {et~al.} 2008, The Astrophysical Journal, 690, 1236, \dodoi{10.1088/0004-637X/690/2/1236}

\bibitem[{{Ivezi{\'c}} {et~al.}(2019){Ivezi{\'c}}, {Kahn}, {Tyson}, {Abel}, {Acosta}, {Allsman}, {Alonso}, {AlSayyad}, {Anderson}, {Andrew}, {Angel}, {Angeli}, {Ansari}, {Antilogus}, {Araujo}, {Armstrong}, {Arndt}, {Astier}, {Aubourg}, {Auza}, {Axelrod}, {Bard}, {Barr}, {Barrau}, {Bartlett}, {Bauer}, {Bauman}, {Baumont}, {Bechtol}, {Bechtol}, {Becker}, {Becla}, {Beldica}, {Bellavia}, {Bianco}, {Biswas}, {Blanc}, {Blazek}, {Blandford}, {Bloom}, {Bogart}, {Bond}, {Booth}, {Borgland}, {Borne}, {Bosch}, {Boutigny}, {Brackett}, {Bradshaw}, {Brandt}, {Brown}, {Bullock}, {Burchat}, {Burke}, {Cagnoli}, {Calabrese}, {Callahan}, {Callen}, {Carlin}, {Carlson}, {Chandrasekharan}, {Charles-Emerson}, {Chesley}, {Cheu}, {Chiang}, {Chiang}, {Chirino}, {Chow}, {Ciardi}, {Claver}, {Cohen-Tanugi}, {Cockrum}, {Coles}, {Connolly}, {Cook}, {Cooray}, {Covey}, {Cribbs}, {Cui}, {Cutri}, {Daly}, {Daniel}, {Daruich}, {Daubard}, {Daues}, {Dawson}, {Delgado}, {Dellapenna}, {de Peyster}, {de Val-Borro}, {Digel}, {Doherty}, {Dubois},
  {Dubois-Felsmann}, {Durech}, {Economou}, {Eifler}, {Eracleous}, {Emmons}, {Fausti Neto}, {Ferguson}, {Figueroa}, {Fisher-Levine}, {Focke}, {Foss}, {Frank}, {Freemon}, {Gangler}, {Gawiser}, {Geary}, {Gee}, {Geha}, {Gessner}, {Gibson}, {Gilmore}, {Glanzman}, {Glick}, {Goldina}, {Goldstein}, {Goodenow}, {Graham}, {Gressler}, {Gris}, {Guy}, {Guyonnet}, {Haller}, {Harris}, {Hascall}, {Haupt}, {Hernandez}, {Herrmann}, {Hileman}, {Hoblitt}, {Hodgson}, {Hogan}, {Howard}, {Huang}, {Huffer}, {Ingraham}, {Innes}, {Jacoby}, {Jain}, {Jammes}, {Jee}, {Jenness}, {Jernigan}, {Jevremovi{\'c}}, {Johns}, {Johnson}, {Johnson}, {Jones}, {Juramy-Gilles}, {Juri{\'c}}, {Kalirai}, {Kallivayalil}, {Kalmbach}, {Kantor}, {Karst}, {Kasliwal}, {Kelly}, {Kessler}, {Kinnison}, {Kirkby}, {Knox}, {Kotov}, {Krabbendam}, {Krughoff}, {Kub{\'a}nek}, {Kuczewski}, {Kulkarni}, {Ku}, {Kurita}, {Lage}, {Lambert}, {Lange}, {Langton}, {Le Guillou}, {Levine}, {Liang}, {Lim}, {Lintott}, {Long}, {Lopez}, {Lotz}, {Lupton}, {Lust}, {MacArthur}, {Mahabal},
  {Mandelbaum}, {Markiewicz}, {Marsh}, {Marshall}, {Marshall}, {May}, {McKercher}, {McQueen}, {Meyers}, {Migliore}, {Miller}, \& {Mills}}]{Ivezic2019}
{Ivezi{\'c}}, {\v{Z}}., {Kahn}, S.~M., {Tyson}, J.~A., {et~al.} 2019, \apj, 873, 111, \dodoi{10.3847/1538-4357/ab042c}

\bibitem[{{Katsianis} {et~al.}(2017){Katsianis}, {Blanc}, {Lagos}, {Tejos}, {Bower}, {Alavi}, {Gonzalez}, {Theuns}, {Schaller}, \& {Lopez}}]{Katsianis2017}
{Katsianis}, A., {Blanc}, G., {Lagos}, C.~P., {et~al.} 2017, \mnras, 472, 919, \dodoi{10.1093/mnras/stx2020}

\bibitem[{{Katsianis} {et~al.}(2019){Katsianis}, {Zheng}, {Gonzalez}, {Blanc}, {Lagos}, {Davies}, {Camps}, {Tr{\v{c}}ka}, {Baes}, {Schaye}, {Trayford}, {Theuns}, \& {Stalevski}}]{Katsianis2019}
{Katsianis}, A., {Zheng}, X., {Gonzalez}, V., {et~al.} 2019, \apj, 879, 11, \dodoi{10.3847/1538-4357/ab1f8d}

\bibitem[{{Katsianis} {et~al.}(2025){Katsianis}, {Wang}, {Yang}, {Zheng}, {Cataldi}, {Napolitano}, {Zhu}, {Tejos}, {Cui}, {Li}, {Lin}, {Feng}, {Li}, {Tang}, {Li}, \& {Pu}}]{Katsianis2025}
{Katsianis}, A., {Wang}, Q., {Yang}, X., {et~al.} 2025, \mnras, 540, 688, \dodoi{10.1093/mnras/staf755}

\bibitem[{{Kravtsov} \& {Borgani}(2012)}]{Kravtsov2012}
{Kravtsov}, A.~V., \& {Borgani}, S. 2012, \araa, 50, 353, \dodoi{10.1146/annurev-astro-081811-125502}

\bibitem[{{Laigle} {et~al.}(2016){Laigle}, {McCracken}, {Ilbert}, {Hsieh}, {Davidzon}, {Capak}, {Hasinger}, {Silverman}, {Pichon}, {Coupon}, {Aussel}, {Le Borgne}, {Caputi}, {Cassata}, {Chang}, {Civano}, {Dunlop}, {Fynbo}, {Kartaltepe}, {Koekemoer}, {Le F{\`e}vre}, {Le Floc'h}, {Leauthaud}, {Lilly}, {Lin}, {Marchesi}, {Milvang-Jensen}, {Salvato}, {Sanders}, {Scoville}, {Smolcic}, {Stockmann}, {Taniguchi}, {Tasca}, {Toft}, {Vaccari}, \& {Zabl}}]{Laigle2016}
{Laigle}, C., {McCracken}, H.~J., {Ilbert}, O., {et~al.} 2016, \apjs, 224, 24, \dodoi{10.3847/0067-0049/224/2/24}

\bibitem[{{Laureijs} {et~al.}(2011){Laureijs}, {Amiaux}, {Arduini}, {Augu{\`e}res}, {Brinchmann}, {Cole}, {Cropper}, {Dabin}, {Duvet}, {Ealet}, {Garilli}, {Gondoin}, {Guzzo}, {Hoar}, {Hoekstra}, {Holmes}, {Kitching}, {Maciaszek}, {Mellier}, {Pasian}, {Percival}, {Rhodes}, {Saavedra Criado}, {Sauvage}, {Scaramella}, {Valenziano}, {Warren}, {Bender}, {Castander}, {Cimatti}, {Le F{\`e}vre}, {Kurki-Suonio}, {Levi}, {Lilje}, {Meylan}, {Nichol}, {Pedersen}, {Popa}, {Rebolo Lopez}, {Rix}, {Rottgering}, {Zeilinger}, {Grupp}, {Hudelot}, {Massey}, {Meneghetti}, {Miller}, {Paltani}, {Paulin-Henriksson}, {Pires}, {Saxton}, {Schrabback}, {Seidel}, {Walsh}, {Aghanim}, {Amendola}, {Bartlett}, {Baccigalupi}, {Beaulieu}, {Benabed}, {Cuby}, {Elbaz}, {Fosalba}, {Gavazzi}, {Helmi}, {Hook}, {Irwin}, {Kneib}, {Kunz}, {Mannucci}, {Moscardini}, {Tao}, {Teyssier}, {Weller}, {Zamorani}, {Zapatero Osorio}, {Boulade}, {Foumond}, {Di Giorgio}, {Guttridge}, {James}, {Kemp}, {Martignac}, {Spencer}, {Walton}, {Bl{\"u}mchen}, {Bonoli},
  {Bortoletto}, {Cerna}, {Corcione}, {Fabron}, {Jahnke}, {Ligori}, {Madrid}, {Martin}, {Morgante}, {Pamplona}, {Prieto}, {Riva}, {Toledo}, {Trifoglio}, {Zerbi}, {Abdalla}, {Douspis}, {Grenet}, {Borgani}, {Bouwens}, {Courbin}, {Delouis}, {Dubath}, {Fontana}, {Frailis}, {Grazian}, {Koppenh{\"o}fer}, {Mansutti}, {Melchior}, {Mignoli}, {Mohr}, {Neissner}, {Noddle}, {Poncet}, {Scodeggio}, {Serrano}, {Shane}, {Starck}, {Surace}, {Taylor}, {Verdoes-Kleijn}, {Vuerli}, {Williams}, {Zacchei}, {Altieri}, {Escudero Sanz}, {Kohley}, {Oosterbroek}, {Astier}, {Bacon}, {Bardelli}, {Baugh}, {Bellagamba}, {Benoist}, {Bianchi}, {Biviano}, {Branchini}, {Carbone}, {Cardone}, {Clements}, {Colombi}, {Conselice}, {Cresci}, {Deacon}, {Dunlop}, {Fedeli}, {Fontanot}, {Franzetti}, {Giocoli}, {Garcia-Bellido}, {Gow}, {Heavens}, {Hewett}, {Heymans}, {Holland}, {Huang}, {Ilbert}, {Joachimi}, {Jennins}, {Kerins}, {Kiessling}, {Kirk}, {Kotak}, {Krause}, {Lahav}, {van Leeuwen}, {Lesgourgues}, {Lombardi}, {Magliocchetti}, {Maguire},
  {Majerotto}, {Maoli}, {Marulli}, {Maurogordato}, {McCracken}, {McLure}, {Melchiorri}, {Merson}, {Moresco}, {Nonino}, {Norberg}, {Peacock}, {Pello}, {Penny}, {Pettorino}, {Di Porto}, {Pozzetti}, {Quercellini}, {Radovich}, {Rassat}, {Roche}, {Ronayette}, \& {Rossetti}}]{Laureijs2011}
{Laureijs}, R., {Amiaux}, J., {Arduini}, S., {et~al.} 2011, arXiv e-prints, arXiv:1110.3193, \dodoi{10.48550/arXiv.1110.3193}

\bibitem[{{Lavoie} {et~al.}(2016){Lavoie}, {Willis}, {D{\'e}mocl{\`e}s}, {Eckert}, {Gastaldello}, {Smith}, {Lidman}, {Adami}, {Pacaud}, {Pierre}, {Clerc}, {Giles}, {Lieu}, {Chiappetti}, {Altieri}, {Ardila}, {Baldry}, {Bongiorno}, {Desai}, {Elyiv}, {Faccioli}, {Gardner}, {Garilli}, {Groote}, {Guennou}, {Guzzo}, {Hopkins}, {Liske}, {McGee}, {Melnyk}, {Owers}, {Poggianti}, {Ponman}, {Scodeggio}, {Spitler}, \& {Tuffs}}]{Lavoie2016}
{Lavoie}, S., {Willis}, J.~P., {D{\'e}mocl{\`e}s}, J., {et~al.} 2016, \mnras, 462, 4141, \dodoi{10.1093/mnras/stw1906}

\bibitem[{{Leja} {et~al.}(2013){Leja}, {van Dokkum}, \& {Franx}}]{Leja2013}
{Leja}, J., {van Dokkum}, P., \& {Franx}, M. 2013, \apj, 766, 33, \dodoi{10.1088/0004-637X/766/1/33}

\bibitem[{{Li} {et~al.}(2022){Li}, {Yang}, {Liu}, {Jing}, {He}, {Huang}, {Dai}, {Sawicki}, {Arnouts}, {Gwyn}, {Moutard}, {Mo}, {Wang}, {Katsianis}, {Cui}, {Han}, {Chiu}, {Gu}, \& {Xu}}]{Li2022}
{Li}, Q., {Yang}, X., {Liu}, C., {et~al.} 2022, \apj, 933, 9, \dodoi{10.3847/1538-4357/ac6e69}

\bibitem[{{Lidman} {et~al.}(2012){Lidman}, {Suherli}, {Muzzin}, {Wilson}, {Demarco}, {Brough}, {Rettura}, {Cox}, {DeGroot}, {Yee}, {Gilbank}, {Hoekstra}, {Balogh}, {Ellingson}, {Hicks}, {Nantais}, {Noble}, {Lacy}, {Surace}, \& {Webb}}]{Lidman2012}
{Lidman}, C., {Suherli}, J., {Muzzin}, A., {et~al.} 2012, \mnras, 427, 550, \dodoi{10.1111/j.1365-2966.2012.21984.x}

\bibitem[{{Lin} {et~al.}(2013){Lin}, {Brodwin}, {Gonzalez}, {Bode}, {Eisenhardt}, {Stanford}, \& {Vikhlinin}}]{Lin2013}
{Lin}, Y.-T., {Brodwin}, M., {Gonzalez}, A.~H., {et~al.} 2013, \apj, 771, 61, \dodoi{10.1088/0004-637X/771/1/61}

\bibitem[{{Lin} \& {Mohr}(2004)}]{Lin2004}
{Lin}, Y.-T., \& {Mohr}, J.~J. 2004, \apj, 617, 879, \dodoi{10.1086/425412}

\bibitem[{{McCracken} {et~al.}(2012){McCracken}, {Milvang-Jensen}, {Dunlop}, {Franx}, {Fynbo}, {Le F{\`e}vre}, {Holt}, {Caputi}, {Goranova}, {Buitrago}, {Emerson}, {Freudling}, {Hudelot}, {L{\'o}pez-Sanjuan}, {Magnard}, {Mellier}, {M{\o}ller}, {Nilsson}, {Sutherland}, {Tasca}, \& {Zabl}}]{McCracken2012}
{McCracken}, H.~J., {Milvang-Jensen}, B., {Dunlop}, J., {et~al.} 2012, \aap, 544, A156, \dodoi{10.1051/0004-6361/201219507}

\bibitem[{{McDonald} {et~al.}(2016){McDonald}, {Stalder}, {Bayliss}, {Allen}, {Applegate}, {Ashby}, {Bautz}, {Benson}, {Bleem}, {Brodwin}, {Carlstrom}, {Chiu}, {Desai}, {Gonzalez}, {Hlavacek-Larrondo}, {Holzapfel}, {Marrone}, {Miller}, {Reichardt}, {Saliwanchik}, {Saro}, {Schrabback}, {Stanford}, {Stark}, {Vieira}, \& {Zenteno}}]{McDonald2016}
{McDonald}, M., {Stalder}, B., {Bayliss}, M., {et~al.} 2016, \apj, 817, 86, \dodoi{10.3847/0004-637X/817/2/86}

\bibitem[{{Mo} {et~al.}(2010){Mo}, {van den Bosch}, \& {White}}]{Galaxybook}
{Mo}, H., {van den Bosch}, F.~C., \& {White}, S. 2010, {Galaxy Formation and Evolution}

\bibitem[{{Moneti} {et~al.}(2023){Moneti}, {McCracken}, {Hudelot}, {Rouberol}, {Herent}, {Mellier}, {Dunlop}, {Le Fevre}, {Franx}, {Fynbo}, {Bowler}, {Caputi}, {Kauffmann}, {Milvang-Jensen}, {Gonzalez-Fernandez}, {Gonzalez-Solares}, {Irwin}, {Lewis}, {Blake}, {Cross}, {Read}, \& {Sutorius}}]{Moneti2023}
{Moneti}, A., {McCracken}, H.~J., {Hudelot}, W., {et~al.} 2023, VizieR Online Data Catalog, II/373

\bibitem[{Morley {et~al.}(2012)Morley, Fortney, Marley, Visscher, Saumon, \& Leggett}]{Morley2012}
Morley, C.~V., Fortney, J.~J., Marley, M.~S., {et~al.} 2012, The Astrophysical Journal, 756, 172, \dodoi{10.1088/0004-637X/756/2/172}

\bibitem[{Morley {et~al.}(2014)Morley, Marley, Fortney, Lupu, Saumon, Greene, \& Lodders}]{Morley2014}
Morley, C.~V., Marley, M.~S., Fortney, J.~J., {et~al.} 2014, The Astrophysical Journal, 787, 78, \dodoi{10.1088/0004-637X/787/1/78}

\bibitem[{{Moster} {et~al.}(2013){Moster}, {Naab}, \& {White}}]{Moster2013}
{Moster}, B.~P., {Naab}, T., \& {White}, S. D.~M. 2013, \mnras, 428, 3121, \dodoi{10.1093/mnras/sts261}

\bibitem[{{Moustakas} {et~al.}(2023){Moustakas}, {Lang}, {Dey}, {Juneau}, {Meisner}, {Myers}, {Schlafly}, {Schlegel}, {Valdes}, {Weaver}, \& {Zhou}}]{Moustakas2023}
{Moustakas}, J., {Lang}, D., {Dey}, A., {et~al.} 2023, \apjs, 269, 3, \dodoi{10.3847/1538-4365/acfaa2}

\bibitem[{Muldrew {et~al.}(2015)Muldrew, Hatch, \& Cooke}]{Muldrew2015}
Muldrew, S.~I., Hatch, N.~A., \& Cooke, E.~A. 2015, Monthly Notices of the Royal Astronomical Society, 452, 2528, \dodoi{10.1093/mnras/stv1449}

\bibitem[{{Navarro} {et~al.}(1997){Navarro}, {Frenk}, \& {White}}]{Navarro1997}
{Navarro}, J.~F., {Frenk}, C.~S., \& {White}, S. D.~M. 1997, \apj, 490, 493, \dodoi{10.1086/304888}

\bibitem[{{Oliva-Altamirano} {et~al.}(2014){Oliva-Altamirano}, {Brough}, {Lidman}, {Couch}, {Hopkins}, {Colless}, {Taylor}, {Robotham}, {Gunawardhana}, {Ponman}, {Baldry}, {Bauer}, {Bland-Hawthorn}, {Cluver}, {Cameron}, {Conselice}, {Driver}, {Edge}, {Graham}, {van Kampen}, {Lara-L{\'o}pez}, {Liske}, {L{\'o}pez-S{\'a}nchez}, {Loveday}, {Mahajan}, {Peacock}, {Phillipps}, {Pimbblet}, \& {Sharp}}]{Oliva2014}
{Oliva-Altamirano}, P., {Brough}, S., {Lidman}, C., {et~al.} 2014, \mnras, 440, 762, \dodoi{10.1093/mnras/stu277}

\bibitem[{{Orellana-Gonz{\'a}lez} {et~al.}(2022){Orellana-Gonz{\'a}lez}, {Cerulo}, {Covone}, {Cheng}, {Leiton}, {Demarco}, \& {Gendron-Marsolais}}]{Orellana2022}
{Orellana-Gonz{\'a}lez}, G., {Cerulo}, P., {Covone}, G., {et~al.} 2022, \mnras, 512, 2758, \dodoi{10.1093/mnras/stac001}

\bibitem[{{Oser} {et~al.}(2010){Oser}, {Ostriker}, {Naab}, {Johansson}, \& {Burkert}}]{Oser2010}
{Oser}, L., {Ostriker}, J.~P., {Naab}, T., {Johansson}, P.~H., \& {Burkert}, A. 2010, \apj, 725, 2312, \dodoi{10.1088/0004-637X/725/2/2312}

\bibitem[{{Overzier}(2016)}]{Overzier2016}
{Overzier}, R.~A. 2016, \aapr, 24, 14, \dodoi{10.1007/s00159-016-0100-3}

\bibitem[{Pickles(1998)}]{Pickles1998}
Pickles, A.~J. 1998, Publications of the Astronomical Society of the Pacific, 110, 863, \dodoi{10.1086/316197}

\bibitem[{{Planck Collaboration} {et~al.}(2020){Planck Collaboration}, {Aghanim}, {Akrami}, {Ashdown}, {Aumont}, {Baccigalupi}, {Ballardini}, {Banday}, {Barreiro}, {Bartolo}, {Basak}, {Battye}, {Benabed}, {Bernard}, {Bersanelli}, {Bielewicz}, {Bock}, {Bond}, {Borrill}, {Bouchet}, {Boulanger}, {Bucher}, {Burigana}, {Butler}, {Calabrese}, {Cardoso}, {Carron}, {Challinor}, {Chiang}, {Chluba}, {Colombo}, {Combet}, {Contreras}, {Crill}, {Cuttaia}, {de Bernardis}, {de Zotti}, {Delabrouille}, {Delouis}, {Di Valentino}, {Diego}, {Dor{\'e}}, {Douspis}, {Ducout}, {Dupac}, {Dusini}, {Efstathiou}, {Elsner}, {En{\ss}lin}, {Eriksen}, {Fantaye}, {Farhang}, {Fergusson}, {Fernandez-Cobos}, {Finelli}, {Forastieri}, {Frailis}, {Fraisse}, {Franceschi}, {Frolov}, {Galeotta}, {Galli}, {Ganga}, {G{\'e}nova-Santos}, {Gerbino}, {Ghosh}, {Gonz{\'a}lez-Nuevo}, {G{\'o}rski}, {Gratton}, {Gruppuso}, {Gudmundsson}, {Hamann}, {Handley}, {Hansen}, {Herranz}, {Hildebrandt}, {Hivon}, {Huang}, {Jaffe}, {Jones}, {Karakci}, {Keih{\"a}nen},
  {Keskitalo}, {Kiiveri}, {Kim}, {Kisner}, {Knox}, {Krachmalnicoff}, {Kunz}, {Kurki-Suonio}, {Lagache}, {Lamarre}, {Lasenby}, {Lattanzi}, {Lawrence}, {Le Jeune}, {Lemos}, {Lesgourgues}, {Levrier}, {Lewis}, {Liguori}, {Lilje}, {Lilley}, {Lindholm}, {L{\'o}pez-Caniego}, {Lubin}, {Ma}, {Mac{\'\i}as-P{\'e}rez}, {Maggio}, {Maino}, {Mandolesi}, {Mangilli}, {Marcos-Caballero}, {Maris}, {Martin}, {Martinelli}, {Mart{\'\i}nez-Gonz{\'a}lez}, {Matarrese}, {Mauri}, {McEwen}, {Meinhold}, {Melchiorri}, {Mennella}, {Migliaccio}, {Millea}, {Mitra}, {Miville-Desch{\^e}nes}, {Molinari}, {Montier}, {Morgante}, {Moss}, {Natoli}, {N{\o}rgaard-Nielsen}, {Pagano}, {Paoletti}, {Partridge}, {Patanchon}, {Peiris}, {Perrotta}, {Pettorino}, {Piacentini}, {Polastri}, {Polenta}, {Puget}, {Rachen}, {Reinecke}, {Remazeilles}, {Renzi}, {Rocha}, {Rosset}, {Roudier}, {Rubi{\~n}o-Mart{\'\i}n}, {Ruiz-Granados}, {Salvati}, {Sandri}, {Savelainen}, {Scott}, {Shellard}, {Sirignano}, {Sirri}, {Spencer}, {Sunyaev}, {Suur-Uski}, {Tauber}, {Tavagnacco},
  {Tenti}, {Toffolatti}, {Tomasi}, {Trombetti}, {Valenziano}, {Valiviita}, {Van Tent}, {Vibert}, {Vielva}, {Villa}, {Vittorio}, {Wandelt}, {Wehus}, {White}, {White}, {Zacchei}, \& {Zonca}}]{Planck2020}
{Planck Collaboration}, {Aghanim}, N., {Akrami}, Y., {et~al.} 2020, \aap, 641, A6, \dodoi{10.1051/0004-6361/201833910}

\bibitem[{{Polletta} {et~al.}(2007){Polletta}, {Tajer}, {Maraschi}, {Trinchieri}, {Lonsdale}, {Chiappetti}, {Andreon}, {Pierre}, {Le F{\`e}vre}, {Zamorani}, {Maccagni}, {Garcet}, {Surdej}, {Franceschini}, {Alloin}, {Shupe}, {Surace}, {Fang}, {Rowan-Robinson}, {Smith}, \& {Tresse}}]{Polletta2007}
{Polletta}, M., {Tajer}, M., {Maraschi}, L., {et~al.} 2007, \apj, 663, 81, \dodoi{10.1086/518113}

\bibitem[{{Ragone-Figueroa} {et~al.}(2018){Ragone-Figueroa}, {Granato}, {Ferraro}, {Murante}, {Biffi}, {Borgani}, {Planelles}, \& {Rasia}}]{Ragone2018}
{Ragone-Figueroa}, C., {Granato}, G.~L., {Ferraro}, M.~E., {et~al.} 2018, \mnras, 479, 1125, \dodoi{10.1093/mnras/sty1639}

\bibitem[{Remus {et~al.}(2023)Remus, Dolag, \& Dannerbauer}]{Remus2023}
Remus, R.-S., Dolag, K., \& Dannerbauer, H. 2023, The Astrophysical Journal, 950, 191, \dodoi{10.3847/1538-4357/accb91}

\bibitem[{Ruszkowski \& Springel(2009)}]{Ruszkowski2009}
Ruszkowski, M., \& Springel, V. 2009, The Astrophysical Journal, 696, 1094, \dodoi{10.1088/0004-637X/696/2/1094}

\bibitem[{Salvato {et~al.}(2008)Salvato, Hasinger, Ilbert, Zamorani, Brusa, Scoville, Rau, Capak, Arnouts, Aussel, Bolzonella, Buongiorno, Cappelluti, Caputi, Civano, Cook, Elvis, Gilli, Jahnke, Kartaltepe, Impey, Lamareille, Floch, Lilly, Mainieri, McCarthy, McCracken, Mignoli, Mobasher, Murayama, Sasaki, Sanders, Schiminovich, Shioya, Shopbell, Silverman, Smolčić, Surace, Taniguchi, Thompson, Trump, Urry, \& Zamojski}]{Salvato2009}
Salvato, M., Hasinger, G., Ilbert, O., {et~al.} 2008, The Astrophysical Journal, 690, 1250, \dodoi{10.1088/0004-637X/690/2/1250}

\bibitem[{{Sawicki} {et~al.}(2020){Sawicki}, {Arcila-Osejo}, {Golob}, {Moutard}, {Arnouts}, \& {Cheema}}]{Sawicki2020}
{Sawicki}, M., {Arcila-Osejo}, L., {Golob}, A., {et~al.} 2020, \mnras, 494, 1366, \dodoi{10.1093/mnras/staa779}

\bibitem[{{Sawicki} {et~al.}(2019){Sawicki}, {Arnouts}, {Huang}, {Coupon}, {Golob}, {Gwyn}, {Foucaud}, {Moutard}, {Iwata}, {Liu}, {Chen}, {Desprez}, {Harikane}, {Ono}, {Strauss}, {Tanaka}, {Thibert}, {Balogh}, {Bundy}, {Chapman}, {Gunn}, {Hsieh}, {Ilbert}, {Jing}, {LeF{\`e}vre}, {Li}, {Matsuda}, {Miyazaki}, {Nagao}, {Nishizawa}, {Ouchi}, {Shimasaku}, {Silverman}, {de la Torre}, {Tresse}, {Wang}, {Willott}, {Yamada}, {Yang}, \& {Yee}}]{Sawicki2019}
{Sawicki}, M., {Arnouts}, S., {Huang}, J., {et~al.} 2019, \mnras, 489, 5202, \dodoi{10.1093/mnras/stz2522}

\bibitem[{{Shankar} {et~al.}(2015){Shankar}, {Buchan}, {Rettura}, {Bouillot}, {Moreno}, {Licitra}, {Bernardi}, {Huertas-Company}, {Mei}, {Ascaso}, {Sheth}, {Delaye}, \& {Raichoor}}]{Shankar2015}
{Shankar}, F., {Buchan}, S., {Rettura}, A., {et~al.} 2015, \apj, 802, 73, \dodoi{10.1088/0004-637X/802/2/73}

\bibitem[{{Sheth} {et~al.}(2001){Sheth}, {Mo}, \& {Tormen}}]{Sheth2001}
{Sheth}, R.~K., {Mo}, H.~J., \& {Tormen}, G. 2001, \mnras, 323, 1, \dodoi{10.1046/j.1365-8711.2001.04006.x}

\bibitem[{{Stott} {et~al.}(2008){Stott}, {Edge}, {Smith}, {Swinbank}, \& {Ebeling}}]{Stott2008}
{Stott}, J.~P., {Edge}, A.~C., {Smith}, G.~P., {Swinbank}, A.~M., \& {Ebeling}, H. 2008, \mnras, 384, 1502, \dodoi{10.1111/j.1365-2966.2007.12807.x}

\bibitem[{{Takada} {et~al.}(2014){Takada}, {Ellis}, {Chiba}, {Greene}, {Aihara}, {Arimoto}, {Bundy}, {Cohen}, {Dor{\'e}}, {Graves}, {Gunn}, {Heckman}, {Hirata}, {Ho}, {Kneib}, {Le F{\`e}vre}, {Lin}, {More}, {Murayama}, {Nagao}, {Ouchi}, {Seiffert}, {Silverman}, {Sodr{\'e}}, {Spergel}, {Strauss}, {Sugai}, {Suto}, {Takami}, \& {Wyse}}]{Takada2014}
{Takada}, M., {Ellis}, R.~S., {Chiba}, M., {et~al.} 2014, \pasj, 66, R1, \dodoi{10.1093/pasj/pst019}

\bibitem[{{Tomczak} {et~al.}(2016){Tomczak}, {Quadri}, {Tran}, {Labb{\'e}}, {Straatman}, {Papovich}, {Glazebrook}, {Allen}, {Brammer}, {Cowley}, {Dickinson}, {Elbaz}, {Inami}, {Kacprzak}, {Morrison}, {Nanayakkara}, {Persson}, {Rees}, {Salmon}, {Schreiber}, {Spitler}, \& {Whitaker}}]{Tomczak2016}
{Tomczak}, A.~R., {Quadri}, R.~F., {Tran}, K.-V.~H., {et~al.} 2016, \apj, 817, 118, \dodoi{10.3847/0004-637X/817/2/118}

\bibitem[{{Tonini} {et~al.}(2012){Tonini}, {Bernyk}, {Croton}, {Maraston}, \& {Thomas}}]{Tonini2012}
{Tonini}, C., {Bernyk}, M., {Croton}, D., {Maraston}, C., \& {Thomas}, D. 2012, \apj, 759, 43, \dodoi{10.1088/0004-637X/759/1/43}

\bibitem[{{van den Bosch} {et~al.}(2004){van den Bosch}, {Norberg}, {Mo}, \& {Yang}}]{van2004}
{van den Bosch}, F.~C., {Norberg}, P., {Mo}, H.~J., \& {Yang}, X. 2004, \mnras, 352, 1302, \dodoi{10.1111/j.1365-2966.2004.08021.x}

\bibitem[{{van Dokkum} {et~al.}(2010){van Dokkum}, {Whitaker}, {Brammer}, {Franx}, {Kriek}, {Labb{\'e}}, {Marchesini}, {Quadri}, {Bezanson}, {Illingworth}, {Muzzin}, {Rudnick}, {Tal}, \& {Wake}}]{vanDokkum2010}
{van Dokkum}, P.~G., {Whitaker}, K.~E., {Brammer}, G., {et~al.} 2010, \apj, 709, 1018, \dodoi{10.1088/0004-637X/709/2/1018}

\bibitem[{Wang {et~al.}(2014)Wang, Mo, Yang, Jing, \& Lin}]{Wang2014}
Wang, H., Mo, H.~J., Yang, X., Jing, Y.~P., \& Lin, W.~P. 2014, The Astrophysical Journal, 794, 94, \dodoi{10.1088/0004-637X/794/1/94}

\bibitem[{{Wang} {et~al.}(2023{\natexlab{a}}){Wang}, {Mo}, {Li}, \& {Chen}}]{Wang2023b}
{Wang}, K., {Mo}, H., {Li}, C., \& {Chen}, Y. 2023{\natexlab{a}}, \mnras, 520, 1774, \dodoi{10.1093/mnras/stad262}

\bibitem[{{Wang} {et~al.}(2023{\natexlab{b}}){Wang}, {Mo}, {Chen}, {Wang}, {Yang}, {Wang}, {Peng}, \& {Cai}}]{Wang2023}
{Wang}, K., {Mo}, H.~J., {Chen}, Y., {et~al.} 2023{\natexlab{b}}, arXiv e-prints, arXiv:2309.01039, \dodoi{10.48550/arXiv.2309.01039}

\bibitem[{{Weaver} {et~al.}(2022){Weaver}, {Kauffmann}, {Ilbert}, {McCracken}, {Moneti}, {Toft}, {Brammer}, {Shuntov}, {Davidzon}, {Hsieh}, {Laigle}, {Anastasiou}, {Jespersen}, {Vinther}, {Capak}, {Casey}, {McPartland}, {Milvang-Jensen}, {Mobasher}, {Sanders}, {Zalesky}, {Arnouts}, {Aussel}, {Dunlop}, {Faisst}, {Franx}, {Furtak}, {Fynbo}, {Gould}, {Greve}, {Gwyn}, {Kartaltepe}, {Kashino}, {Koekemoer}, {Kokorev}, {Le F{\`e}vre}, {Lilly}, {Masters}, {Magdis}, {Mehta}, {Peng}, {Riechers}, {Salvato}, {Sawicki}, {Scarlata}, {Scoville}, {Shirley}, {Silverman}, {Sneppen}, {Smolc̆i{\'c}}, {Steinhardt}, {Stern}, {Tanaka}, {Taniguchi}, {Teplitz}, {Vaccari}, {Wang}, \& {Zamorani}}]{Weaver2022}
{Weaver}, J.~R., {Kauffmann}, O.~B., {Ilbert}, O., {et~al.} 2022, \apjs, 258, 11, \dodoi{10.3847/1538-4365/ac3078}

\bibitem[{Weaver {et~al.}(2023)Weaver, Zalesky, Kokorev, McPartland, Chartab, Gould, Shuntov, Davidzon, Faisst, Stickley, Capak, Toft, Masters, Mobasher, Sanders, Kauffmann, McCracken, Ilbert, Brammer, \& Moneti}]{Weaver2023}
Weaver, J.~R., Zalesky, L., Kokorev, V., {et~al.} 2023, The Astrophysical Journal Supplement Series, 269, 20, \dodoi{10.3847/1538-4365/acf850}

\bibitem[{{Webb} {et~al.}(2015){Webb}, {Muzzin}, {Noble}, {Bonaventura}, {Geach}, {Hezaveh}, {Lidman}, {Wilson}, {Yee}, {Surace}, \& {Shupe}}]{Webb2015}
{Webb}, T. M.~A., {Muzzin}, A., {Noble}, A., {et~al.} 2015, \apj, 814, 96, \dodoi{10.1088/0004-637X/814/2/96}

\bibitem[{{Wen} \& {Han}(2011)}]{Wen2011}
{Wen}, Z.~L., \& {Han}, J.~L. 2011, \apj, 734, 68, \dodoi{10.1088/0004-637X/734/1/68}

\bibitem[{{Whiley} {et~al.}(2008){Whiley}, {Arag{\'o}n-Salamanca}, {De Lucia}, {von der Linden}, {Bamford}, {Best}, {Bremer}, {Jablonka}, {Johnson}, {Milvang-Jensen}, {Noll}, {Poggianti}, {Rudnick}, {Saglia}, {White}, \& {Zaritsky}}]{Whiley2008}
{Whiley}, I.~M., {Arag{\'o}n-Salamanca}, A., {De Lucia}, G., {et~al.} 2008, \mnras, 387, 1253, \dodoi{10.1111/j.1365-2966.2008.13324.x}

\bibitem[{{Williams} {et~al.}(2024){Williams}, {Damjanov}, {Sawicki}, {Souchereau}, {Chen}, {Desprez}, {George}, {Annunziatella}, {Arnouts}, {Gwyn}, {Marchesini}, \& {Sajina}}]{Williams2024}
{Williams}, D.~J., {Damjanov}, I., {Sawicki}, M., {et~al.} 2024, arXiv e-prints, arXiv:2412.03662, \dodoi{10.48550/arXiv.2412.03662}

\bibitem[{{Willmer}(2018)}]{Willmer2018}
{Willmer}, C. N.~A. 2018, \apjs, 236, 47, \dodoi{10.3847/1538-4365/aabfdf}

\bibitem[{{Yang} {et~al.}(2003){Yang}, {Mo}, \& {van den Bosch}}]{Yang2003}
{Yang}, X., {Mo}, H.~J., \& {van den Bosch}, F.~C. 2003, \mnras, 339, 1057, \dodoi{10.1046/j.1365-8711.2003.06254.x}

\bibitem[{{Yang} {et~al.}(2013){Yang}, {Mo}, {van den Bosch}, {Bonaca}, {Li}, {Lu}, {Lu}, \& {Lu}}]{Yang2013}
{Yang}, X., {Mo}, H.~J., {van den Bosch}, F.~C., {et~al.} 2013, \apj, 770, 115, \dodoi{10.1088/0004-637X/770/2/115}

\bibitem[{{Yang} {et~al.}(2005){Yang}, {Mo}, {van den Bosch}, \& {Jing}}]{Yang2005}
{Yang}, X., {Mo}, H.~J., {van den Bosch}, F.~C., \& {Jing}, Y.~P. 2005, \mnras, 356, 1293, \dodoi{10.1111/j.1365-2966.2005.08560.x}

\bibitem[{{Yang} {et~al.}(2007){Yang}, {Mo}, {van den Bosch}, {Pasquali}, {Li}, \& {Barden}}]{Yang2007}
{Yang}, X., {Mo}, H.~J., {van den Bosch}, F.~C., {et~al.} 2007, \apj, 671, 153, \dodoi{10.1086/522027}

\bibitem[{{Yang} {et~al.}(2012){Yang}, {Mo}, {van den Bosch}, {Zhang}, \& {Han}}]{Yang2012}
{Yang}, X., {Mo}, H.~J., {van den Bosch}, F.~C., {Zhang}, Y., \& {Han}, J. 2012, \apj, 752, 41, \dodoi{10.1088/0004-637X/752/1/41}

\bibitem[{{Yang} {et~al.}(2021){Yang}, {Xu}, {He}, {Gu}, {Katsianis}, {Meng}, {Shi}, {Zou}, {Zhang}, {Liu}, {Wang}, {Dong}, {Lu}, {Li}, {Chen}, {Wang}, {Mo}, {Fu}, {Guo}, {Leauthaud}, {Luo}, {Zhang}, \& {Zu}}]{Yang2021}
{Yang}, X., {Xu}, H., {He}, M., {et~al.} 2021, \apj, 909, 143, \dodoi{10.3847/1538-4357/abddb2}

\bibitem[{{Zhan}(2011)}]{Zhan2011}
{Zhan}, H. 2011, Scientia Sinica Physica, Mechanica \& Astronomica, 41, 1441, \dodoi{10.1360/132011-961}

\bibitem[{{Zhang} {et~al.}(2016){Zhang}, {Miller}, {McKay}, {Rooney}, {Evrard}, {Romer}, {Perfecto}, {Song}, {Desai}, {Mohr}, {Wilcox}, {Bermeo-Hernandez}, {Jeltema}, {Hollowood}, {Bacon}, {Capozzi}, {Collins}, {Das}, {Gerdes}, {Hennig}, {Hilton}, {Hoyle}, {Kay}, {Liddle}, {Mann}, {Mehrtens}, {Nichol}, {Papovich}, {Sahl{\'e}n}, {Soares-Santos}, {Stott}, {Viana}, {Abbott}, {Abdalla}, {Banerji}, {Bauer}, {Benoit-L{\'e}vy}, {Bertin}, {Brooks}, {Buckley-Geer}, {Burke}, {Carnero Rosell}, {Castander}, {Diehl}, {Doel}, {Cunha}, {Eifler}, {Fausti Neto}, {Fernandez}, {Flaugher}, {Fosalba}, {Frieman}, {Gaztanaga}, {Gruen}, {Gruendl}, {Honscheid}, {James}, {Kuehn}, {Kuropatkin}, {Lahav}, {Maia}, {Makler}, {Marshall}, {Martini}, {Miquel}, {Ogando}, {Plazas}, {Roodman}, {Rykoff}, {Sako}, {Sanchez}, {Scarpine}, {Schubnell}, {Sevilla}, {Smith}, {Sobreira}, {Suchyta}, {Swanson}, {Tarle}, {Thaler}, {Tucker}, {Vikram}, \& {da Costa}}]{Zhang2016}
{Zhang}, Y., {Miller}, C., {McKay}, T., {et~al.} 2016, \apj, 816, 98, \dodoi{10.3847/0004-637X/816/2/98}

\bibitem[{{Zhao} {et~al.}(2017){Zhao}, {Conselice}, {Arag{\'o}n-Salamanca}, {Almaini}, {Hartley}, {Lani}, {Mortlock}, \& {Old}}]{Zhao2017}
{Zhao}, D., {Conselice}, C.~J., {Arag{\'o}n-Salamanca}, A., {et~al.} 2017, \mnras, 464, 1393, \dodoi{10.1093/mnras/stw2406}

\end{thebibliography}
\bibliographystyle{aasjournal}

\end{document}